\documentclass[12pt,preprint]{aastex}

\newcommand{\kms}{km\,s$^{-1}$}
\newcommand{\kkms}{K\,km\,s$^{-1}$}
\newcommand{\percc}{cm$^{-3}$}
\newcommand{\percms}{$\mathrm{cm^{-2}}$}
\newcommand{\percmsmag}{$\mathrm{cm^{-2}}$~$\mathrm{mag^{-1}}$}
\newcommand{\htwo}{H$_2$}
\newcommand{\twco}{$^{12}$CO}
\newcommand{\thco}{$^{13}$CO}
\newcommand{\ceo}{C$^{18}$O}
\newcommand{\kmspc}{km\,s$^{-1}$\,pc$^{-1}$}
\newcommand{\tastar}{$T^*_A$}
\newcommand{\trstar}{$T^*_R$}
\newcommand{\hthcop}{H$^{13}$CO$^+$}
\newcommand{\dcop}{DCO$^+$}
\newcommand{\av}{$A_V$}

\slugcomment{To appear in ApJS}

\shorttitle{Lupus Molecular Clouds}
\shortauthors{Tothill et al.}

\begin{document}

\title{Large-Scale CO Maps of the Lupus Molecular Cloud Complex}

\author{N.~F.~H.~Tothill\altaffilmark{1,2}, A. L\"ohr\altaffilmark{1}, 
S.~C.~Parshley\altaffilmark{3}, A.~A.~Stark\altaffilmark{1}, 
A.~P.~Lane\altaffilmark{1}, J.~I.~Harnett\altaffilmark{1,4},
G.~A.~Wright\altaffilmark{5}, C.~K.~Walker\altaffilmark{6}, 
T.~L.~Bourke\altaffilmark{1} and P.~C.~Myers\altaffilmark{1}}




\altaffiltext{1}{Harvard-Smithsonian Center for Astrophysics, 
60 Garden Street, Cambridge, MA 02138, USA}
\altaffiltext{2}{School of Physics, University of Exeter,
Stocker Road, Exeter, EX\,4~4\,QL, UK; nfht@astro.ex.ac.uk}
\altaffiltext{3}{Department of Astronomy, Cornell University, Ithaca, 
NY 14853, USA}
\altaffiltext{4}{National Radio Astronomy Observatory, Green Bank, WV, USA}
\altaffiltext{5}{Antiope Associates, 18 Clay Street, Fair Haven, 
NJ 07704, USA}
\altaffiltext{6}{Steward Observatory, University of Arizona, Tucson, 
AZ 85721, USA}

\begin{abstract}
Fully sampled degree-scale maps of the $^{13}$CO 2--1 and CO 4--3 
transitions toward three members of the Lupus Molecular Cloud Complex --- 
Lupus~I, III, and IV --- trace the column density and temperature of the 
molecular gas. Comparison with IR
extinction maps from the c2d project requires most of the gas to have a
temperature of 8--10\,K. Estimates of the cloud mass from $^{13}$CO
emission are roughly consistent with most previous estimates, while the 
line widths are higher, around 2~km\,s$^{-1}$. CO 4--3 emission is found 
throughout Lupus~I, indicating widespread dense gas, and toward Lupus~III 
and IV. Enhanced line widths at the NW end and along the edge of the 
B\,228 ridge in Lupus~I, and a coherent velocity gradient across the ridge,
are consistent with interaction between the molecular cloud and an expanding 
\ion{H}{1} shell from the Upper-Scorpius subgroup of the Sco-Cen OB 
Association. Lupus~III is dominated by the effects of two HAe/Be stars, and 
shows no sign of external influence. Slightly warmer gas around the core of 
Lupus~IV and a low line width suggest heating by the Upper-Centaurus-Lupus 
subgroup of Sco-Cen, without the effects of an \ion{H}{1} shell.
\end{abstract}



\section{Introduction}

The Lupus star forming region, recently reviewed by \citet{comeron}, lies 
about 150~pc from the Earth \citep{08alombardi}, in the Gould Belt. It is 
immediately visible by inspection of optical photographs of the sky, 
comprising a set of largely filamentary dark clouds. Based on optical 
extinction maps \citep{cambresy}, the `Cores to Disks' Spitzer Legacy 
Programme \citep[c2d,][]{c2d,c2dproducts,c2dresults}, has produced and 
analyzed infrared images of three areas: Lupus~I, III, and 
IV\footnote{Some authors use arabic numerals: Lupus~1, 3 and 4}. 
The c2d data products \citep{c2dproducts} include: shorter-wavelength 
Infrared Array Camera (IRAC) maps, used to identify and classify young 
stellar objects \citep[YSOs,][]{merin}; far-IR MIPS maps \citep{mips} 
tracing the thermal emission of the dust component of the molecular clouds; 
and IR extinction maps, based on the Two Micron All Sky 
Survey \citep[2MASS;][]{mips} and on a combination of 2MASS and IRAC
data \citep{c2dproducts}. In this paper, we present maps of the c2d fields 
in Lupus, tracing the molecular hydrogen by the rotational emission of carbon 
monoxide (CO) and isotopically substituted \thco. 

The Lupus complex has been extensively surveyed in the various 1--0 
transitions of CO: \citet{86murphy} first mapped the Lupus molecular clouds 
using the 1.2~m Columbia telescope with 0.5\degr\ resolution to show the 
extent of the molecular gas in Lupus and suggest a total mass of a few 
$10^4$~$M_\sun$. Improved maps in several transitions were obtained by 
NANTEN (2.7\arcmin\ resolution, but routinely undersampled): \citet{01tachi} 
published \twco\ maps of the whole complex, mostly sampled at 8\arcmin\ 
spacing, with some areas at 4\arcmin\ spacing; \thco\ 1--0 maps 
\citep{96tachi} cover Lupus~I and III with 8\arcmin\ spacing, while \ceo\ 
1--0 maps \citep{99hara} cover all the clouds in the complex at 2\arcmin\ 
spacing. \citet{02my} mapped about 200 arcmin$^2$ of Lupus~IV at 
sub-arcminute resolution and full-beam spacing 
in the 1--0 transitions of CO, \thco, and \ceo. The \thco\ 2--1 and CO 4--3 
maps presented here are the first large-scale fully sampled low-$J$ CO maps 
of Lupus, and the first mid-$J$ CO maps of any kind.

The large-scale structure of the Lupus clouds can be traced by near-IR 
extinction \citep{08blombardi} and CO 1--0 emission \citep{01tachi}. 
The complex covers some 20\degr\ of galactic latitude (about 50~pc). 
At low latitudes ($b\la 10^\circ$), a large mass of diffuse gas contains
denser filamentary clouds; at higher latitudes, the dense clouds (Lupus I 
and II) are more clearly separated. The evolution of the Lupus clouds may
have been driven by the influence of nearby OB stars \citep{01tachi}.
The Lupus and Ophiuchus molecular clouds face one another across the 
Upper-Scorpius subgroup of the Scorpius-Centaurus OB Association; the 
Upper-Centaurus-Lupus subgroup lies on the opposite side of the Lupus
clouds from Upper-Sco. The \ion{H}{1} shell around Upper-Sco, blown by 
stellar winds and a presumed supernova 1.5\,Myr ago \citep{92degeus}, 
borders the NE side of the Lupus clouds; on the plane of the sky, the 
ridge which dominates Lupus~I (B\,228, see below) lies just on the trailing
edge of the \ion{H}{1} shell. The much older Upper-Cen-Lup subgroup has 
driven an \ion{H}{1} shell far beyond the Lupus clouds \citep{92degeus}; it 
should have passed through the Lupus complex some 4--7 Myr ago, roughly 
consistent with the ages of T~Tauri stars in Lupus III and IV \citep{02my}.

\begin{figure}
\includegraphics[scale=0.5]{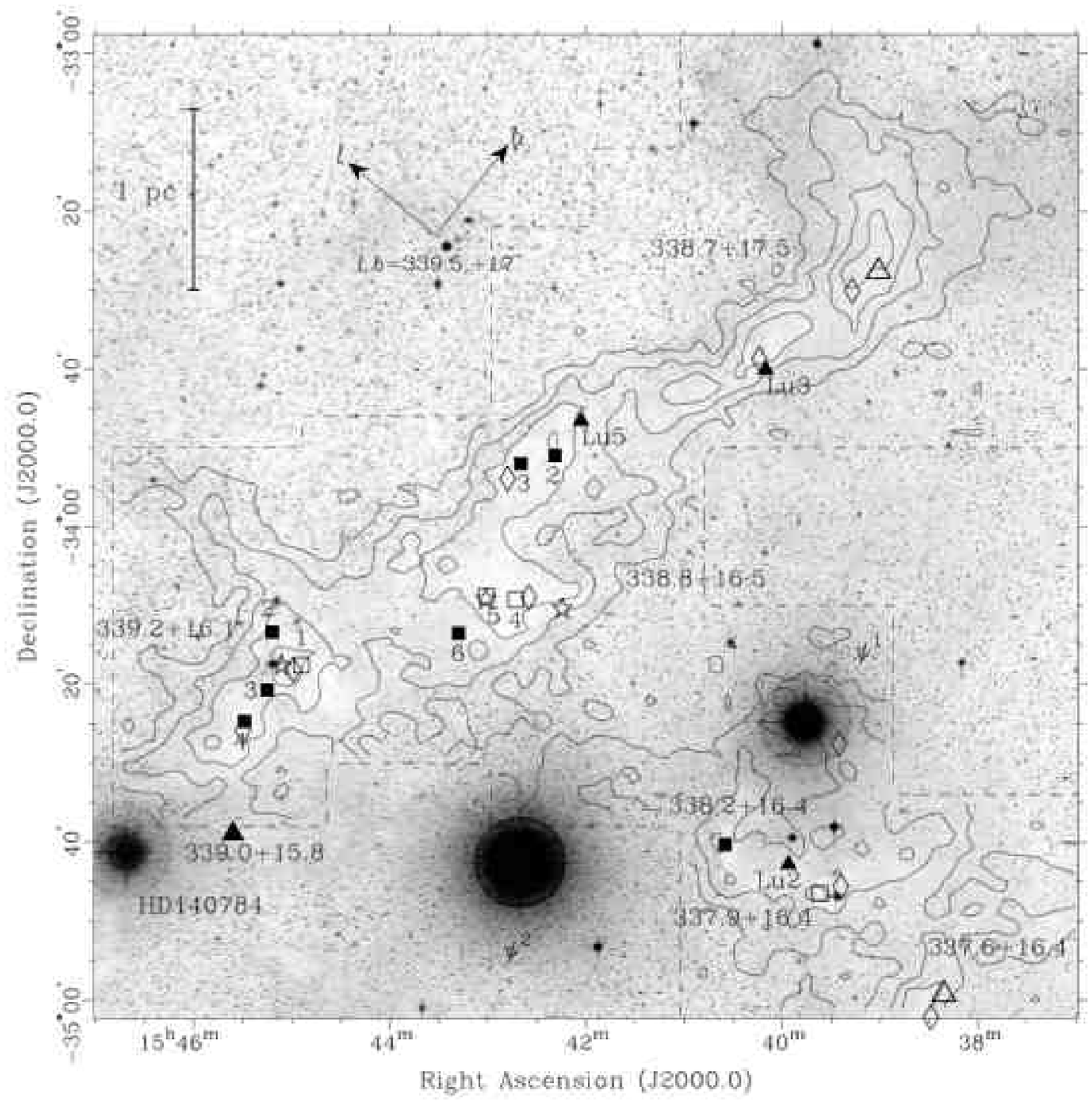}
\caption{Optical (blue DSS2) image of Lupus~I, overlaid with 
\thco\ 2--1 emission from this work (contours at 1, 3, 5, 7, 9\,\kkms).
Annotations: dark cloud positions (Table~\ref{tab-dcs}) from 
\citet[open triangles]{86hartley}, \citet[filled triangles]{vmf},
and \citet[squares]{99lm}; embedded YSOs from \citet[star-shapes]{merin}; 
bright stars (by name); \ceo\ 1--0 peaks from \citet[diamonds]{99hara}. 
Filled squares denote cloud cores observed in molecular lines by 
\citet{04lmp}. Only objects within the \thco\ map are plotted.
The $l,b$ arrows are 15\arcmin\ long.}
\label{fig-lui-opt}
\end{figure}

\begin{figure}
\includegraphics[scale=0.5]{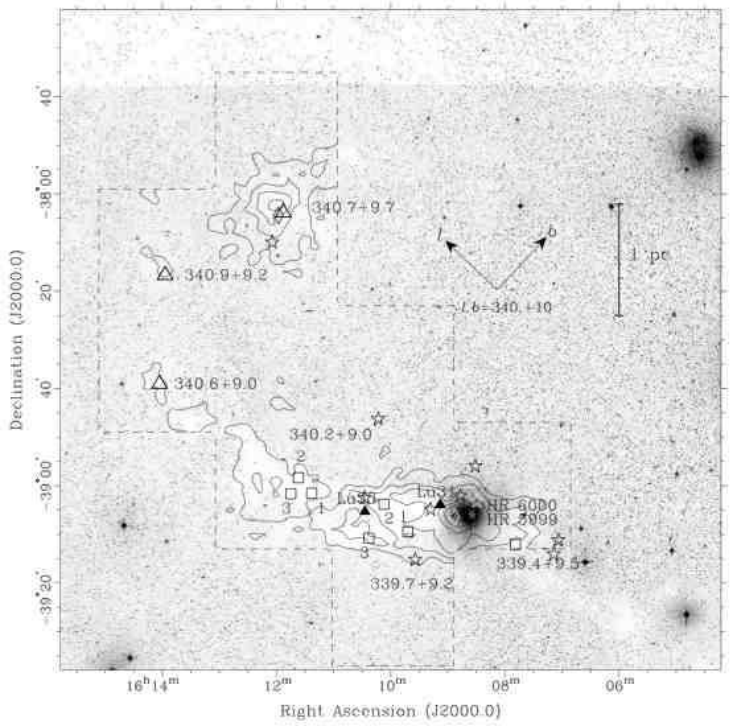}
\caption{Optical image of Lupus~III, as in Figure~\ref{fig-lui-opt}.
\thco\ contours are at 2, 4, \dots 12\,\kkms. Bright stars are marked
as crosses.}
\label{fig-luiii-opt}
\end{figure}

\begin{figure}
\includegraphics[scale=0.5]{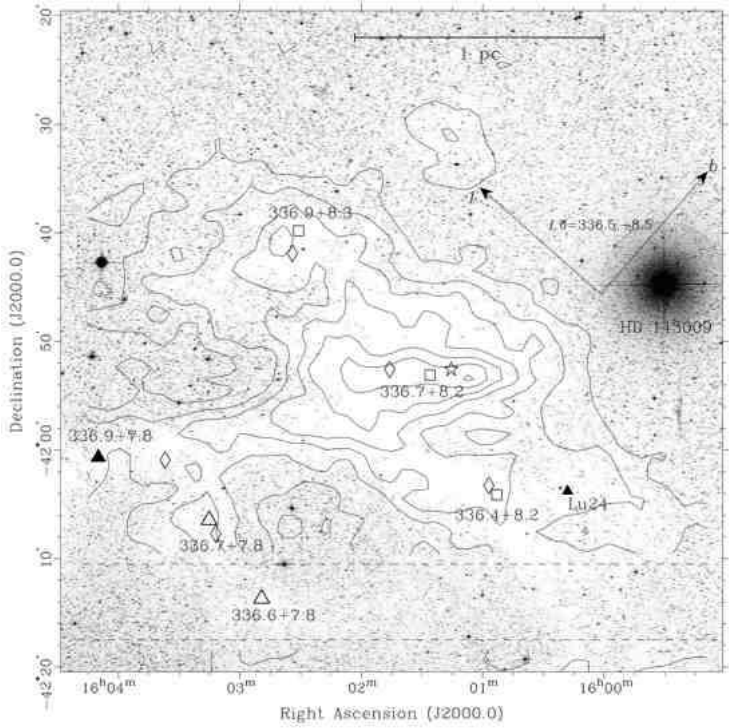}
\caption{Optical image of Lupus~IV, as in Figure~\ref{fig-lui-opt}.
\thco\ contours are at 1, 2, \dots 7\,\kkms.}
\label{fig-luiv-opt}
\end{figure}

Lupus~I (Figure~\ref{fig-lui-opt}) is dominated by B\,228 \citep{barnard}
or GF\,19 \citep{79schneider}, a long ridge running NW--SE, extending over 
about 2\degr\ (5~pc) parallel to the edge of the Upper-Sco \ion{H}{1} shell. 
The molecular material appears to fall off steeply toward the center of 
the shell (at the NE edge), but there is extensive material on the other 
side of the ridge. Lupus~III (Figure~\ref{fig-luiii-opt}) is a long 
(about $4\times 1$~pc) E--W cloud \citep[GF\,21,][]{79schneider} at the 
edge of the low-latitude cloud mass: at its eastern end, it curves up to 
the NE and breaks up into clumps; toward the west lie two embedded (but 
optically visible) Herbig Ae/Be stars, HR\,5999 (A\,5--7) and the 
B\,6 HR\,6000 \citep{comeron}. \citet{02tachi} consider Lupus~III to be 
a cluster-forming structure with a dense head and spread-out tail, 
analogous to the $\rho$~Oph and Cha~I clouds. A couple of pc north 
of Lupus~III, there is another smaller cloud, which appears to be a 
separate condensation in the low-latitude gas; we henceforth refer to 
this northern structure as Lupus~III\,N. Lupus~IV (Figure~\ref{fig-luiv-opt}) 
is the head of a filamentary structure running approximately E--W 
\citep[GF\,17,][]{79schneider,02my}, comprising a small core, about 0.5~pc 
long (E--W), lying within a more diffuse extended cloud, about a pc across. 
The filament connects Lupus~IV to the low-latitude gas mass to the East;
\citeauthor{79schneider} describe it as a ``chain of small faint globules''.

Individual dark clouds or cloud peaks have been identified by inspection 
of optical surveys on scales ranging from degrees (e.g., B\,228) to 
arcminutes \citep{sandy,feitzinger,86hartley,bhr1,96avb,99lm,vmf}; these 
clouds are listed in Table~\ref{tab-dcs}. Near-IR extinction maps of 
Lupus~III with a resolution of 30\arcsec\ \citep{05teix} have been used 
to identify dark cores in more detail, which were then fitted with 
Bonnor-Ebert profiles. Single-pointing spectra in several molecular 
lines have been observed toward extinction-selected clumps: 
\citet{sandy} observed H$_2$CO along four lines of sight at 6~cm 
(6.6\arcmin\ beamwidth); \citet{bhr2} searched for ammonia, but did not 
detect it, toward two globules (BHR\,120 \& 140); \citet{vmf} observed 
\thco\ and \ceo\ 1--0, 
but their 0.8\arcmin\ beamwidth is poorly-matched to ours; \citet{04lmp} 
observed a few of their Lupus~I cores \citep{99lm} in CS 3--2 and 
\dcop\ 2--1 (0.7\arcmin\ beamwidth). Previously identified CO cores, 
both those observed by \citeauthor{vmf}, and those identified from 
their \ceo\ maps by \citet{99hara}, are listed in Table~\ref{tab-mcs}. 
The positions of these extinction cores and \ceo\ cores with respect to 
optical and \thco\ 2--1 emission are shown in 
Figures~\ref{fig-lui-opt}--\ref{fig-luiv-opt}. Of the members of the c2d small
dark cloud sample mapped by AST/RO in \thco\ 2--1 and CO 4--3 
\citep{andrea}, three lie within the Lupus region, but outside our maps.

Lupus~I contains a few YSOs, concentrated to the B\,228 ridge \citep{merin}. 
Lupus~III contains a very dense cluster of YSOs near the Herbig Ae/Be stars, 
composed mainly of T~Tauri stars \citep{07allen,merin}, but including the 
Class~0 object Lupus\,3\,MMS \citep{07tachi}. \citet{09comeron} identified 
a population of cool stars and brown dwarfs toward Lupus~I and III, which 
are not strongly concentrated toward the molecular clouds. The Lupus~IV
region contains about as many YSOs as the Lupus~I region \citep{merin},
but they are older (Class II/III) and largely found outside the molecular 
cloud, with the exception of one flat-spectrum YSO candidate close to the
cloud center. Even the dispersed population is absent \citep{09comeron}.
Lupus~IV thus appears to have less (or no) ongoing star formation, compared
to Lupus~I and III. \citeauthor{merin} suggest that the Class II and III 
YSOs surrounding it represent an earlier generation of star formation 
(possibly associated with the passage of the Upper-Cen-Lup \ion{H}{1} 
shell mentioned above), and that the very dense Lupus~IV core is poised to 
form new stars. The younger YSOs  
\citep[Class 0, I, and flat-spectrum sources from the c2d samples,][]{merin} 
within the boundaries of our \thco\ 2--1 maps are listed in 
Table~\ref{tab-yso}, and plotted in 
Figures~\ref{fig-lui-opt}--\ref{fig-luiv-opt}.

\section{Observations}


Lupus I, III, and IV were observed with the Antarctic Submillimeter Telescope 
and Remote Observatory \citep[AST/RO,][]{obspaper}, during the period 2005 
March--November, in the 461\,GHz CO 4--3 and 220\,GHz \thco\ 2--1 transitions. 
The telescope, receiver, and spectrometer systems are described by 
\citet{obspaper}: \thco\ 2--1 observations used the 230\,GHz SIS receiver, 
whose local oscillator system has been upgraded to use a synthesizer, 
frequency multiplied ($\times 19$) by a Millitech multiplier. CO 4--3 
observations used the 450--495\,GHz SIS waveguide receiver \citep{wanda}. 
Low-resolution acousto-optical spectrometers (AOSs, 1\,GHz bandwidth, 0.7\,MHz 
channel width) were used for all observations.

The large-scale maps presented in this paper were built up from multiple 
smaller maps (`submaps'), mosaicked together as follows: each submap was 
centered on a point of a large-scale grid superimposed on the molecular cloud, 
the grid points separated by 24\arcmin\ and 12\arcmin\ for the 2--1 and 4--3 
observations respectively. The submaps measure $28\arcmin\times 29\arcmin$ 
($\alpha\times\delta$) and $12\arcmin\times 12\arcmin$ respectively, giving 
substantial overlap between neighbouring 2--1 maps and some overlap (one 
row or column of spectra) between neighbouring 4--3 maps.

The submaps use cell sizes of 1\arcmin\ and 0.5\arcmin, significantly 
oversampling the telescope beam (3.3\arcmin\ at 220\,GHz, 1.7\arcmin\ at 
461\,GHz). The spectra are position switched, taken in batches of 4 or 5, 
all of which share a reference measurement, about a degree away. The 4 or 5 
source spectra are all at the same declination (and hence at the same 
elevation, since AST/RO lies at the South Geographic Pole), but the 
reference position is the same for all spectra in a submap, and may thus 
differ in elevation from the source spectra by as much as a few arcminutes. 
The use of one reference position for each submap, however, allows the 
reference position to be checked for emission. Each row of each submap is 
built up from a number of contiguous blocks of reference-sharing spectra. 
Reference sharing increases the observing efficiency almost to that of 
on-the-fly mapping, but also shares one of its weaknesses: the sharing of a 
reference spectrum produces correlated noise in neighboring spectra, which 
can show up as artifacts in the map. Rather than the long stripes 
characteristic of on-the-fly mapping, correlated noise is manifested in the 
AST/RO maps as short (4- or 5-cell) horizontal blocks.

\section{Data Reduction}

The AST/RO observing system produces spectra calibrated onto the 
\tastar\ scale \citep{obspaper}; because of the unobstructed off-axis 
optical design of the telescope, this is essentially equivalent to 
\trstar. Further reduction and processing were carried out (using COMB, 
the standard AST/RO data reduction 
program\footnote{http://www.astro.umd.edu/$\sim$mpound/comb}, and 
PDL\footnote{the Perl Data Language, http://pdl.perl.org}). 
The spectra have velocity resolutions of 0.9\,\kms\ (\thco\ 2--1)
and 0.4\,\kms\ (CO 4--3); they were
checked for frequency shifts, and corrections were applied (corrections
ranged from 0 to almost 5\,\kms\ --- see 
Appendix~\ref{app-data}). Linear baselines were subtracted from all 
spectra, along with polynomial baselines if necessary: a few percent of 
the 4--3 spectra required non-linear baseline subtraction, and almost 
none of the 2--1 spectra (Table~\ref{tab-spectra}).

Long-integration spectra were taken toward all reference positions, 
generally showing upper limits of $\sim 0.1$\,K (\thco\ 2--1) and 
$\sim 0.3$~K (CO 4--3). Some 2--1 reference positions (for one submap 
in Lupus~I and one in Lupus~III) contained substantial emission 
($\sim 1$\,K), and these reference spectra were added back into all the 
spectra in the relevant submap. 

\begin{figure}
\includegraphics[scale=0.7]{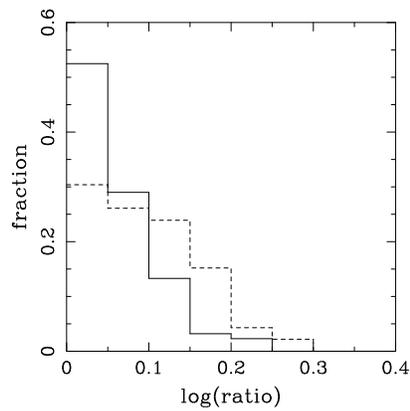}
\caption{Distributions of the logarithms (base 10) of the peak intensity 
ratios for points with multiple \thco\ 2--1 (solid) and CO 4--3 (dashed) 
spectra, in all maps. Only ratios with estimated S/N$>$2 are included.}
\label{fig-peakratio}
\end{figure}

Because of the overlap between submaps, there are over a thousand 
pointings on the sky toward which two or more spectra were measured at 
different times. These duplicate pointings can be used to estimate the 
internal consistency of the flux calibration at the telescope. 
Distributions of the ratios of peak \trstar\ between spectra with the 
same pointings are shown in Figure~\ref{fig-peakratio} for both 4--3 and 
2--1 data. The typical inconsistencies between these spectra (estimated 
by the standard deviation of the distributions of \trstar\ ratios) are about 
30\% for CO 4--3 data, and about 20\% for \thco\ 2--1.

\subsection{Map Making}

Initial map making was carried out by nearest-neighbor sampling the submap 
data onto one large grid, with overlapping observations co-added. Each pixel of 
these maps corresponds directly to the spectrum observed by the telescope; 
without smoothing, inconsistencies between neighbouring submaps show up as 
clear edges, which are much more obvious in the 4--3 than in the 2--1 data.

The overlap between submaps allowed these inconsistencies in the 4--3 data 
to be checked and corrected. Individual submaps with significantly different 
flux scales to their surroundings were identified and rescaled. These 
rescalings were checked visually: a submap was only rescaled if it was 
obviously higher or lower than its neighbours, if the measured scaling was 
consistent with the visual one, and if the rescaling improved the appearance 
of the map when it was applied. Rescaling corrections of $\la 10$\% did not 
significantly improve the appearance of the map, and were not implemented. 
Fewer than half the submaps needed correction, mostly by factors of 0.7--1.4 
(in line with the estimate of internal consistency given above), and the 
largest correction factor was $\sim 1.8$. No corrections needed to be applied 
to the 2--1 data. The need to rescale probably arises from imperfect 
estimation of the atmospheric transparency in the calibration procedure,
which works better when the atmosphere is fairly optically thin 
(e.g.~at 220~GHz) than at higher opacities (e.g., at 461~GHz).

\begin{figure}
\includegraphics[scale=0.9]{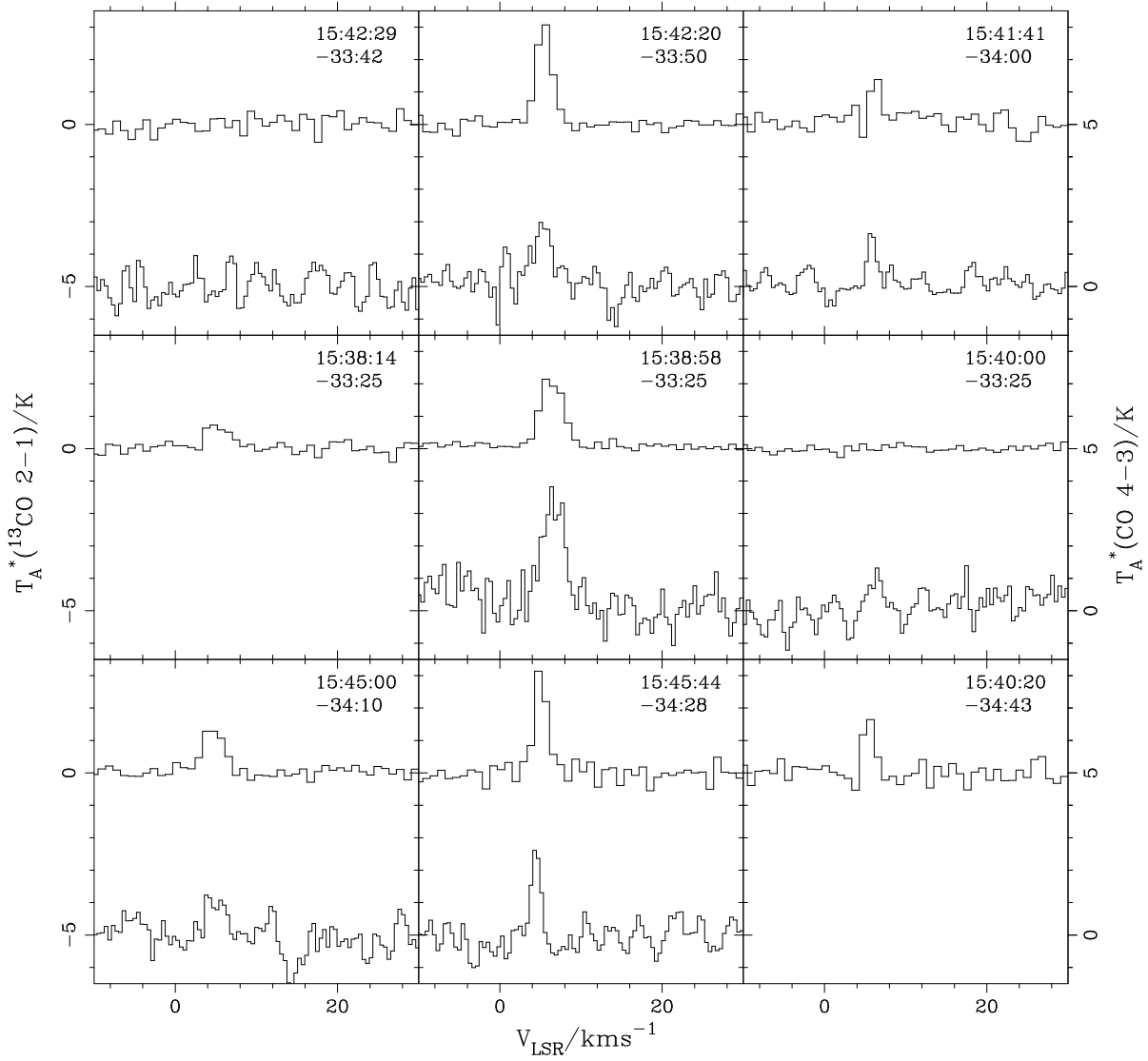}
\caption{Sample spectra toward Lupus~I: each panel shows a \thco~2--1
spectrum (with position annotated). If there is a CO~4--3 spectrum toward 
that position (within 0.25\arcmin), it is shown below the \thco~2--1 spectrum.}
\label{fig-spectra-lui}
\end{figure}

\begin{figure}
\includegraphics[scale=0.9]{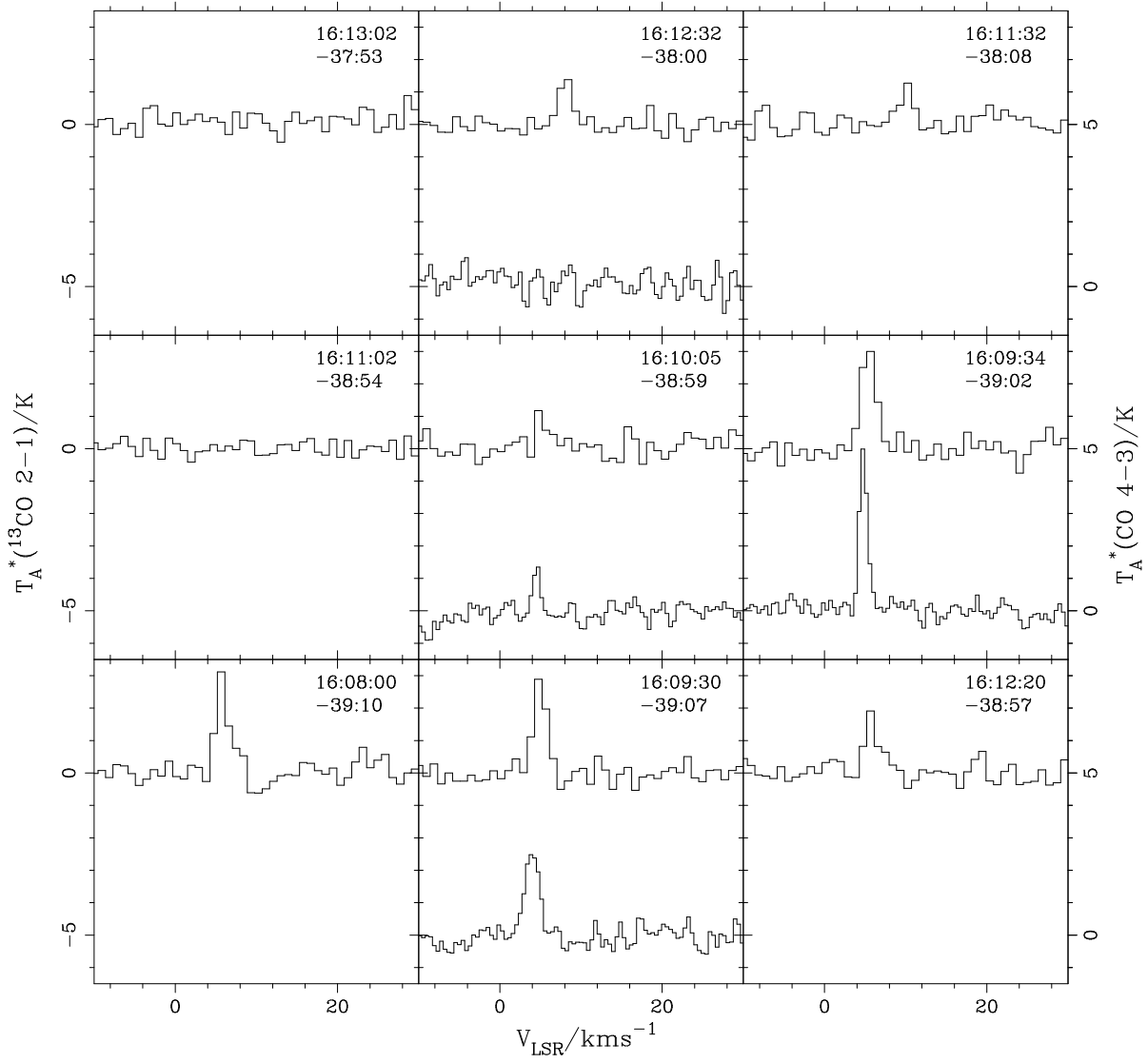}
\caption{Sample spectra toward Lupus~III\,N (upper row) and 
Lupus~III (lower row). See Figure~\ref{fig-spectra-lui} for details.}
\label{fig-spectra-luiii}
\end{figure}

\begin{figure}
\includegraphics[scale=0.9]{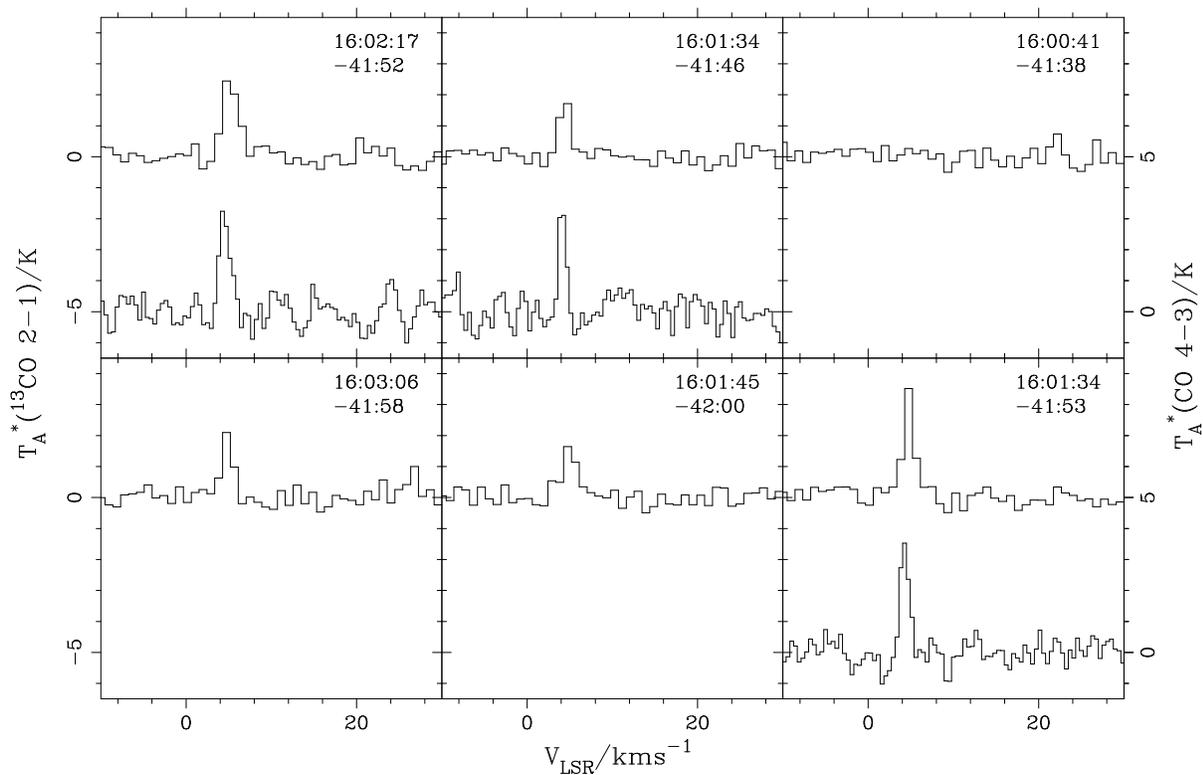}
\caption{Sample spectra toward Lupus~IV. See 
Figure~\ref{fig-spectra-lui} for details.}
\label{fig-spectra-luiv}
\end{figure}

The corrected 4--3 data (and uncorrected 2--1 data) were then gridded (as 
above) to produce final unsmoothed 
data cubes\footnote{publicly available in FITS format at 
\anchor{http://www.astro.ex.ac.uk/people/nfht/resources_lupus.html}
{http://www.astro.ex.ac.uk/people/nfht/resources\_lupus.html}}, 
which were used to generate the maps in this paper. All the quantities 
mapped are calculated directly from the data, without profile-fitting. For 
example, the line width is estimated by the ratio of integrated intensity to 
peak \trstar\ (which overestimates the FWHM by about 10\% if applied to a 
perfect gaussian line). Sample spectra from the data cubes are shown 
in Figures~\ref{fig-spectra-lui}--\ref{fig-spectra-luiv}.

\begin{figure}
\includegraphics[scale=0.7]{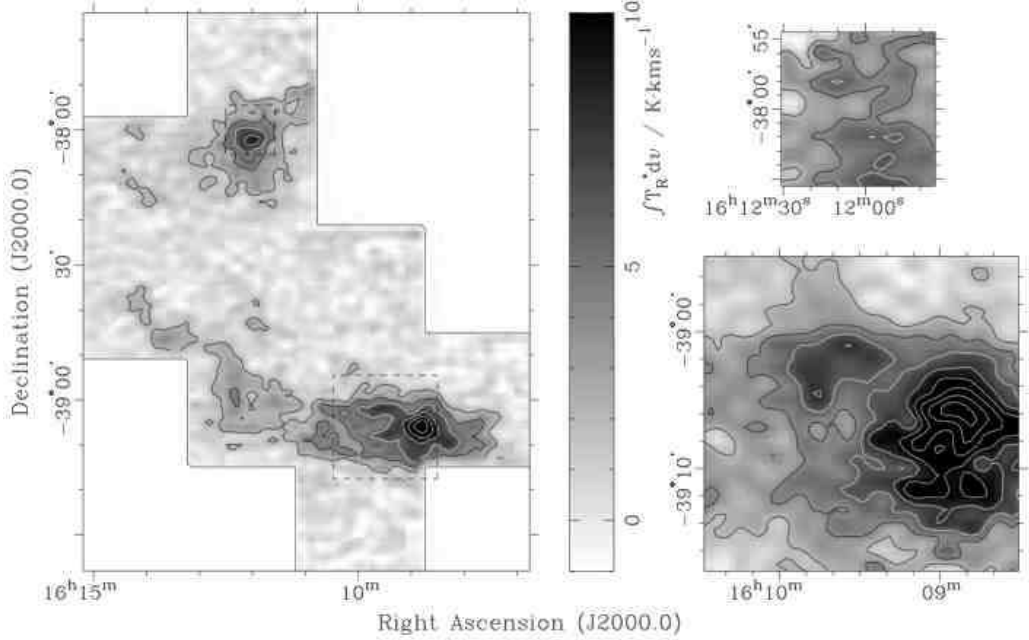}
\caption{Maps of Lupus~III and Lupus~III\,N in (left) \thco~2--1, 
integrated from $v_{LSR}=+2$ to $+12$\,\kms, and (right) CO~4--3,
integrated from $v_{LSR}=+2$ to $+8$\,\kms\ toward Lupus~III and 
from $+6$ to $+12$\,\kms\ toward Lupus~III\,N. Contours at 
2, 4\,\kkms\ (black) and 6, 8\dots 16\,\kkms\ (white).}
\label{fig-luiii-int}
\end{figure}

\begin{figure}
\includegraphics[scale=0.7]{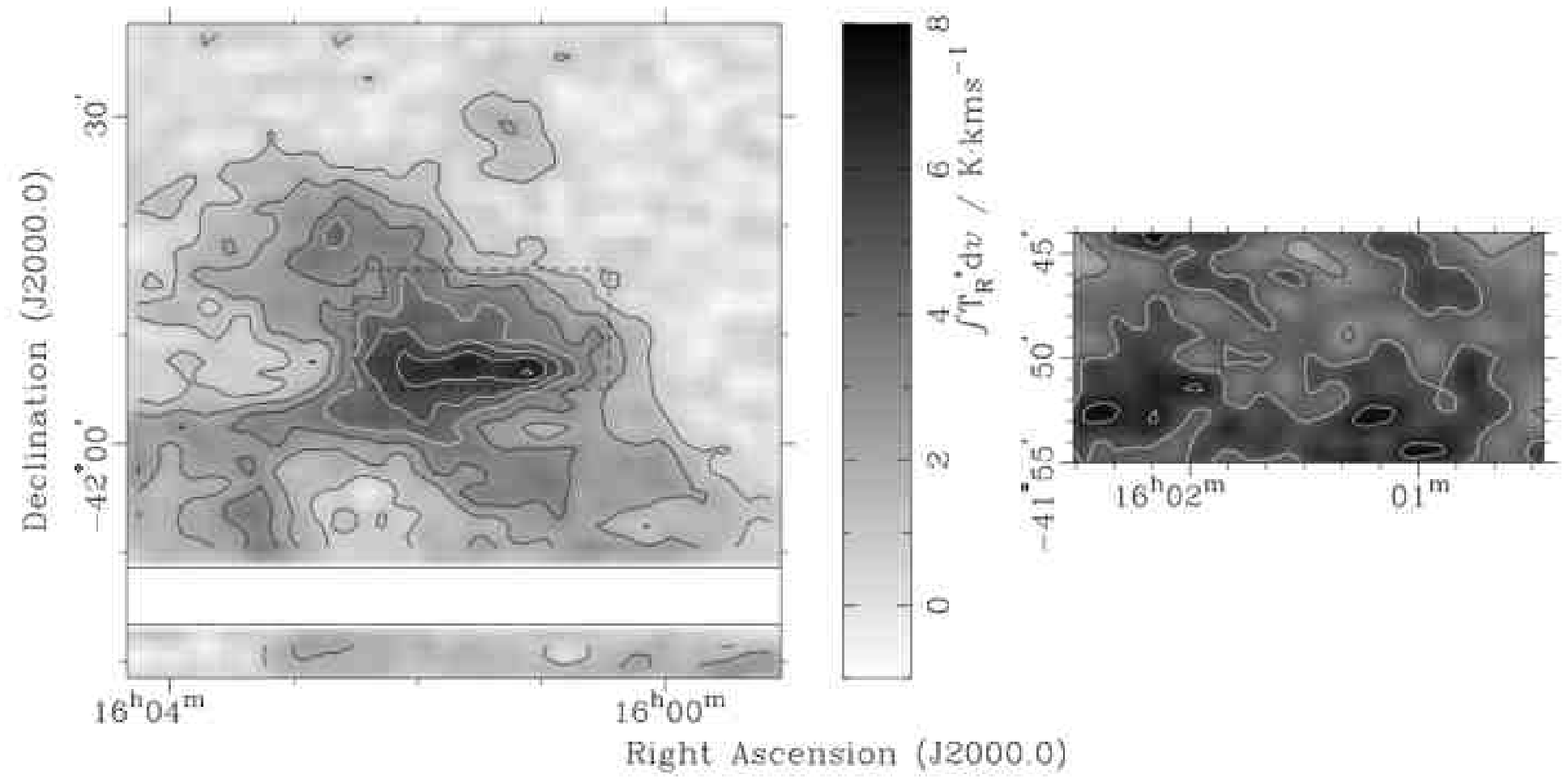}
\caption{Maps of Lupus~IV in (left) \thco~2--1 and (right) CO~4--3,
integrated from $v_{LSR}=+3$ to $+7$\,\kms. \thco~2--1 contours at 
1, 2, 3\,\kkms\ (black) and 4, 5, 6, 7\,\kkms\ (white); CO~4--3 contours 
at 3\,\kkms\ (black) and 5, 7\,\kkms\ (white).}
\label{fig-luiv-int}
\end{figure}

\begin{figure}
\includegraphics[scale=0.85]{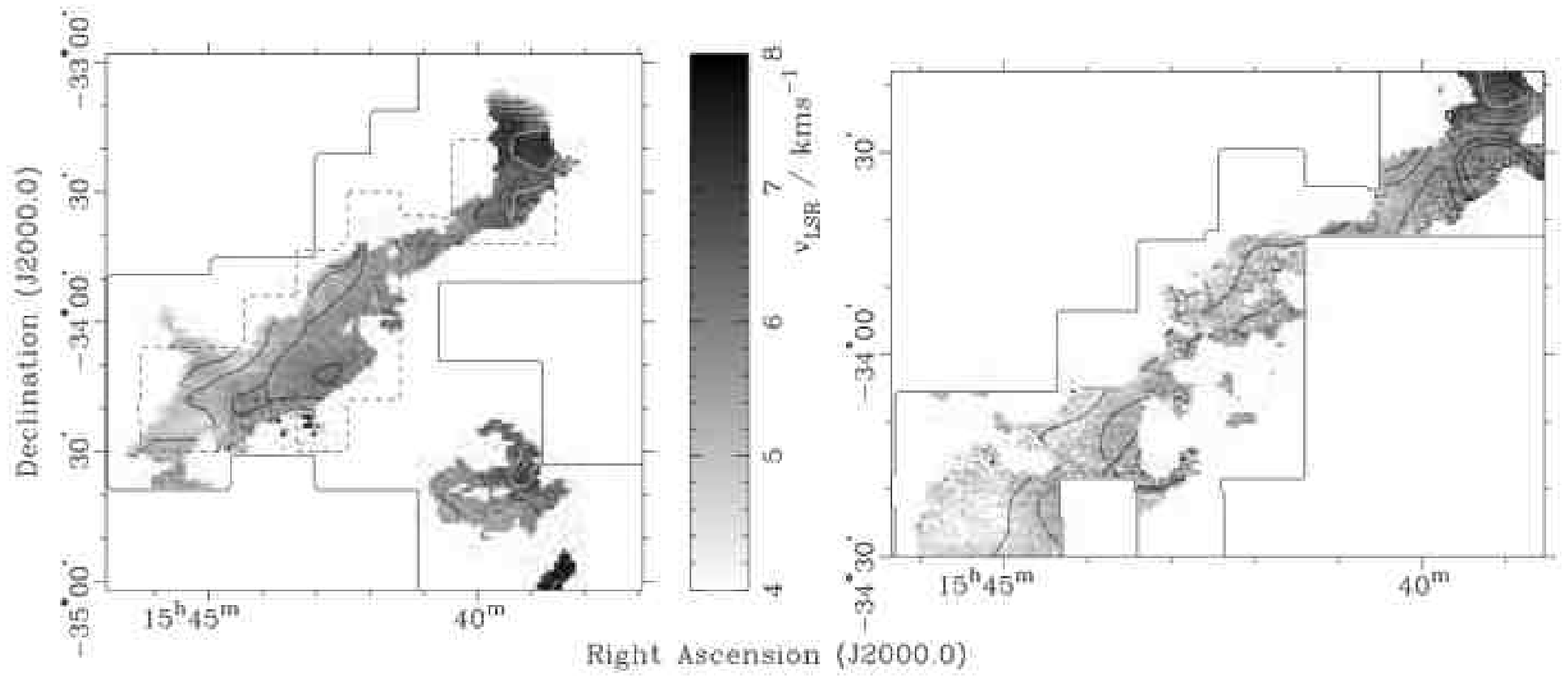}
\caption{Maps of the centroid velocity of (left) \thco\ 2--1 and (right) 
CO 4--3 emission toward Lupus~I. Contours are at 5, 5.5, 6\,\kms\ (black) 
and 6.5, 7, 7.5\,\kms\ (white). The gray scale is unsmoothed, while the 
contours have been smoothed with a Gaussian kernel. Only spectra with 
S/N$>$10 are shown. Solid outlines indicate the extent of the maps, while 
the broken outline indicates the extent of the CO~4--3 map.}
\label{fig-lui-vcen}
\end{figure}

\begin{figure}
\includegraphics[scale=0.75]{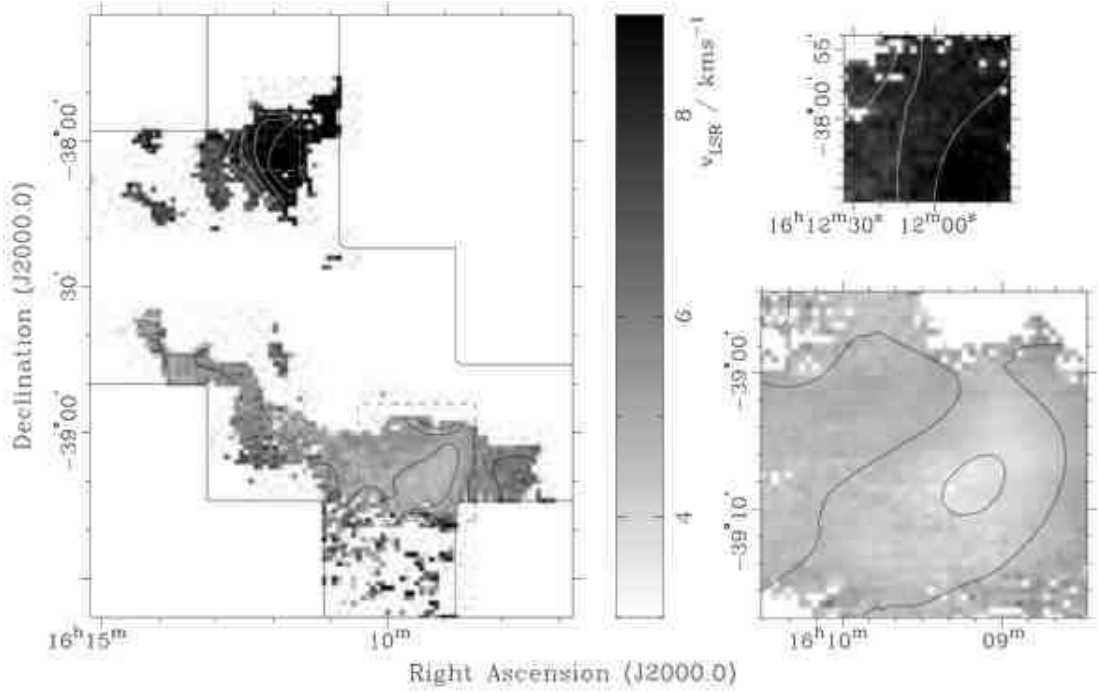}
\caption{Maps of the centroid velocity of emission toward Lupus~III, as in 
Figure~\ref{fig-lui-vcen}: \thco~2--1 contours are at 5, 5.5, 6\,\kms\ 
(black) and 7.5, 8, 8.5, 9\,\kms\ (white); CO~4--3 contours are at
4, 4.5,\kms\ (black) and 7.5, 8, 8.5, 9\,\kms\ (white). Only spectra with 
S/N$>$5 are shown.}
\label{fig-luiii-vcen}
\end{figure}

\begin{figure}
\includegraphics[scale=0.7]{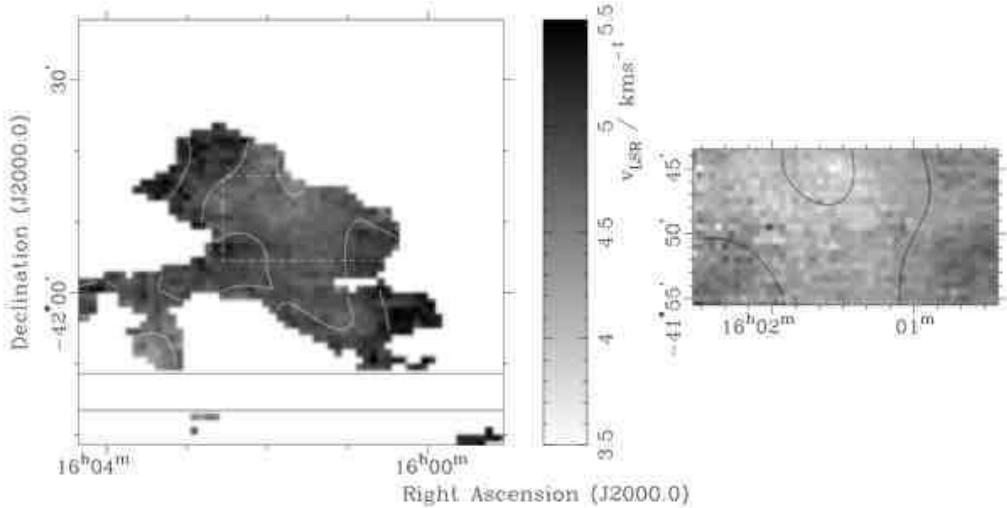}
\caption{Maps of the centroid velocity of emission toward Lupus~IV, as in 
Figure~\ref{fig-lui-vcen}: contours are at 4, 4.3 (black) and 
4.6, 4.9, 5.2\,\kms\ (white). Only spectra with S/N$>$10 are shown.}
\label{fig-luiv-vcen}
\end{figure}

Integrated intensity maps (Figures~\ref{fig-lui-int}--\ref{fig-luiv-int}), 
channel maps (integrated over 2~\kms\ channels, 
Figures~\ref{fig-lui-chan}--\ref{fig-luiv-chan}), and peak \trstar\ maps 
(Figures~\ref{fig-lui-tmax}--\ref{fig-luiv-tmax}) were all produced by 
bessel smoothing \citep[e.g.][]{surf}, which gives the `most fair' 
representation of the sky as 
observed by a single-dish telescope. However, if significant emission extends 
to the edge of the map, this technique tends to produce a roll-off of emission 
in the outermost pixels of the map, and this effect is visible in some of the 
Lupus maps. The effective telescope diameter was picked to balance resolution
against noise: 1.5\,m for the 2--1 maps, and 1.2\,m for the 4--3, yielding 
overall beam sizes of 3.8\arcmin\ and 2.3\arcmin, respectively.
Maps of the velocity centroid (Figures~\ref{fig-lui-vcen}--\ref{fig-luiv-vcen}) 
and line width (Figures~\ref{fig-lui-lwid}--\ref{fig-luiv-lwid}) use both 
unsmoothed data (in grayscale) and Gaussian-smoothed data (contours), with 
Gaussian FWHMs of 5\arcmin\ (2--1) and 5.75\arcmin\ (4--3) yielding 
overall beams of 6\arcmin\ in both transitions.

The nominal pointing accuracy of AST/RO is about 1\arcmin, but comparison of 
the \thco\ 2--1 map of Lupus~IV with the optical extinction structure 
(see~Figure~\ref{fig-luiv-opt}) suggests a systematic offset in declination. 
Discrepancies of similar magnitude can be seen in other AST/RO data from 
2005 \citep{andrea}. The Lupus~IV maps have therefore all been shifted 
southwards by 1.5\arcmin. The submillimeter map is well correlated with 
the millimeter-wave one, and so it has been shifted as well. There is no 
evidence of similar systematic shifts in the Lupus~I or Lupus~III data. The 
publicly available data cubes {\em do not} have this shift applied to them.

\section{Results}

\subsection{Overview of the Lupus Clouds}
\label{sec-results-overview}

\citet{vmf} estimated \ceo\ 1--0 optical depths in the range 0.1--0.5 
toward extinction peaks, implying that \thco\ 2--1 will no longer be 
optically thin in dense parts of the Lupus clouds. Maps of \thco\ 2--1 
peak \trstar\ follow the extinction maps of \citet{mips} better than 
maps of integrated intensity do, suggesting widespread breakdown of the 
optically thin approximation. The maximum peak \trstar\ toward Lupus~I 
and IV is about 3.5\,K in each map, implying a minimum excitation 
temperature, $T_x$, of 8\,K; the peak \trstar\ in Lupus~III reaches 5\,K 
near the bright stars, so $T_x\ga 10$\,K in this area. \citet{96tachi} 
adopted 13\,K for their analysis of 
\thco\ 1--0 toward Lupus~I. A comparison of the lower \thco\ contours 
to the most recent c2d extinction maps suggests that excitation 
temperatures of 8--10~K will reproduce the extinction-traced column 
density quite well. While higher excitation temperatures are not ruled 
out by the molecular line data, they would lead to lower column density 
estimates which would be less compatible with estimates from extinction, 
which is probably the  least biased estimator of the true column 
density \citep{08goodman}.

Gas masses are estimated from the \thco\ 2--1 data as follows: the ratio 
of the peak \trstar\ to the maximum possible brightness temperature (for 
assumed $T_x$) yields an estimate of the peak optical depth 
$\tau_0^{(13)}$; this is multiplied by the line width to estimate 
$\int\tau dv$, which is proportional to the column density of \thco\ 
($N^{(13)}$, Appendix~\ref{app-co}). $N^{(13)}$ is converted to a total 
gas column density by assuming conversions between $N^{(13)}$ and 
extinction $A_V$ (see below), and between $A_V$ and the column density 
of hydrogen 
\citep[$N_{\rm H}=1.37\times 10^{21}A_V$\,\percms,][]{c2dresults}. 
Assuming a distance of 150\,pc, the total gas mass enclosed in a 
1\,arcmin$^2$ pixel is then calculated.

The relationship between \thco\ column density and extinction can be 
expressed as $N^{(13)}=x^{(13)}\times 10^{15}(A_V-A_{V,0})$~cm$^{-2}$, with 
the extinction threshold $A_{V,0}$ reflecting the minimum column density 
required for the presence of \thco. The most commonly used parameters 
come from a study of $\rho$~Oph by 
\citet[][FLW]{flw}: $x^{(13)}=2.7$\,\percmsmag, $A_{V,0}=1.6$\,mag; other 
studies \citep[e.g.][]{bachiller,lada} have found values of $x^{(13)}$ in 
the range 2.2--2.7, and extinction thresholds of 0--1.6~mag. Recently, a 
comprehensive study of the Perseus complex \citep{08pineda} found 
$x^{(13)}=2.4$\,\percmsmag\, $A_{V,0}=1.7$\,mag overall, and 
$x^{(13)}=1.9$\,\percmsmag, $A_{V,0}=1.2$\,mag toward the West End of the 
complex (PWE), comprising several dark clouds, which is likely to be the 
best analog to the Lupus clouds.

Estimated masses of the Lupus clouds above $A_V$ thresholds of 2 and 
3~mag are given in Table~\ref{tab-mass}, along with the mean and 
dispersion of the line widths of material with $A_V\ge 2$. (The line width 
distribution for $A_V\ge 3$ is very similar, varying only by about 
0.1\,\kms.) Masses have been calculated for $T_x=8,9,10$\,K and for 
three \thco-to-$A_V$ relations: PWE, Perseus, and FLW. The quoted mass 
is the estimate using $T_x=9$\,K and PWE, and the positive and negative 
errors give the full range of mass estimates. Other estimates of mass 
and line width from the literature are also tabulated. Previous 
isotopically substituted CO results relied, directly or indirectly, on 
the older ratio of $N_H$ to $A_V$ 
\citep[$1.87\times 10^{21}$\,\percms,][]{bsd}, and have been rescaled 
to use the same ratio as in this work. The extinction mass of Lupus~III 
has also been rescaled to assume a distance of 150\,pc instead of the 
200\,pc used by \citet{mips} and \citet{c2dresults}.

The masses of Lupus~I, III (including Lupus~III\,N) and IV are estimated
to be 280, 150, and 80~$M_\odot$ respectively (for $A_V\ge 2$). The 
extinction mass of Lupus~I (251~$M_\odot$) is similar to the \thco\ 2--1 
estimate; those of Lupus~III and IV are significantly greater (250 and 
120~$M_\odot$), but still consistent with the \thco\ 2--1 error range.
This suggests that the Lupus complex has spatial variations in the 
\thco-to-\av\ ratio, as Perseus does \citep{08goodman}. By contrast,
the \thco\ 1--0 mass estimate toward Lupus~I is 880~$M_\odot$ 
\citep{96tachi}. The \thco\ 1--0 map covers a much larger area than ours
(almost 10\,deg$^2$) and includes significant emission to the 
SE and SW of our map boundaries. Within the confines of the \thco\
2--1 map, we estimate that the \thco\ 1--0 mass should be 
600--700~$M_\odot$, still about 50\% more than the upper limit of our 
mass range. A similar discrepancy was 
reported by \citet{02my}, whose mass estimate for Lupus~II is about a 
factor of 3 lower than that of \citeauthor{96tachi}; but their mass 
estimate for Lupus~IV is quite similar to ours. Possible reasons for the 
discrepancies include the undersampling of the NANTEN maps (which will tend 
to overestimate the area of structures) and differences in the calibration 
of the two telescopes. \citet{99hara} measured only 240\,$M_\odot$ toward 
Lupus~I in \ceo\ 1--0, which is much closer to our estimate. The 
additional mass in the \thco\ 1--0 map might comprise tenuous gas in which 
the 2--1 transition is not fully excited, and whose column density is low 
enough for it not to be detectable in \ceo\ 1--0. The \thco\ 1--0 mass 
estimate toward Lupus~III (220~$M_\odot$) is consistent with our data and 
close to the extinction result, while the \ceo\ 1--0 estimates toward 
Lupus~III and III\,N (80 and 5~$M_\odot$) are much lower.

The average line widths for the Lupus 
clouds \citep[Table~\ref{tab-mass} and][]{c2dresults}\footnote{The 
tabulated line widths are slightly different from those in the earlier 
paper: \citeauthor{c2dresults} averaged the mean and median line widths 
and took the difference between mean and median to be the error range; 
we only report the mean.} have been deconvolved by subtracting the AOS 
channel width in quadrature, and, at 1.5--2.7~\kms, are broader than 
the \thco\ 1--0 widths reported at emission peaks by 
\citet{96tachi}, much broader than the 0.9~\kms\ given by
\citet{99hara} as the typical line width, and also broader than the 
\ceo\ 1--0 line widths toward individual cores (Table~\ref{tab-mcs}). 
\ceo\ 1--0 is expected to display narrower lines than \thco\ 2--1,
since the more optically thin transition arises mainly from the denser
gas in the center of the cores, which generally has a lower velocity
dispersion. It is harder to see why there should be a discrepancy between
\thco\ 1--0 and 2--1 line widths. However, the 1--0 measurements
are taken from two spectra extracted from the map, so the data are 
insufficient to investigate the disagreement.
Despite these differences, the various line width measurements follow 
the same pattern: Lupus~I and III have quite similar line widths, 
Lupus~III\,N has broader lines and Lupus~IV narrower. 

Based on the line widths of \thco\ 2--1 and the velocity gradients 
observed toward the clouds, it is clear that they are gravitationally
unbound (although their general filamentary shape makes the standard
formula for virial mass hard to apply). Lupus~IV, however, may be 
marginally bound due to its high column density, narrow lines and low
velocity gradients: the lower limit to its virial mass is estimated to 
be $\sim$140\,$M_\odot$, while the upper limit to the mass derived from
the \thco\ 2--1 map is $\sim$150\,$M_\odot$. \citet{99hara} found 
almost all of the dense cores that they identified in \ceo\ 1--0
to be gravitationally unbound; the greater line widths in \thco\ 2--1
reinforce this conclusion.

\subsubsection{CO 4--3 Emission}

Because of their small areal coverage, concentrating on \thco\ 2--1 
emission peaks, the CO 4--3 maps do not provide the same extensive 
overview of the clouds. The 4--3 map of Lupus~I, however, covers most of 
the B\,228 ridge. Throughout most of the ridge, the peak \trstar\ lies 
between 1 and 2\,K, corresponding to excitation temperatures of 7--9\,K 
(assuming that the CO emission is optically thick throughout, with a 
filling factor of unity). The ends of the ridge may be warmer, with peak 
\trstar\ of 3--4\,K implying excitation temperatures of 10--12\,K.

The 4--3 maps of Lupus~III and III\,N show the majority of the gas to be 
similar to the ridge in Lupus~I, with $T_x\sim 8-10$\,K. Within the main 
Lupus~III cloud there are two maxima: one with peak \trstar\ of 4\,K 
($T_x\sim 12$\,K), while the other peaks at 9\,K. This latter one appears 
very compact, and so its excitation temperature may be significantly 
higher than the 18\,K minimum implied by its peak \trstar. The 4--3 map of 
Lupus~IV covers only the very center of the cloud, with peak \trstar\ of 
3--4\,K ($T_x\sim$\,10--12\,K).

\subsection{Lupus I}

\begin{figure}
\includegraphics[scale=0.7]{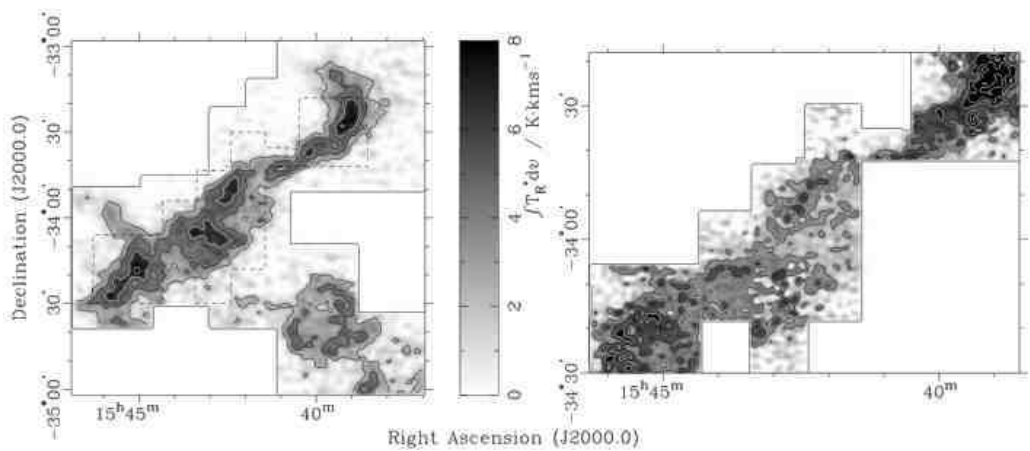}
\caption{Maps of Lupus~I in (left) \thco~2--1 and (right) CO~4--3,
integrated from $v_{LSR}=+2$ to $+10$\,\kms. \thco~2--1 contours at 
2, 4\,\kkms\ (black) and 6, 8\,\kkms\ (white); CO~4--3 contours 
at 3, 5 (black) and 7, 9, 11\,\kkms\ (white). The broken outline denotes
the extent of the CO~4--3 map.}
\label{fig-lui-int}
\end{figure}

\begin{figure}
\includegraphics[scale=0.8]{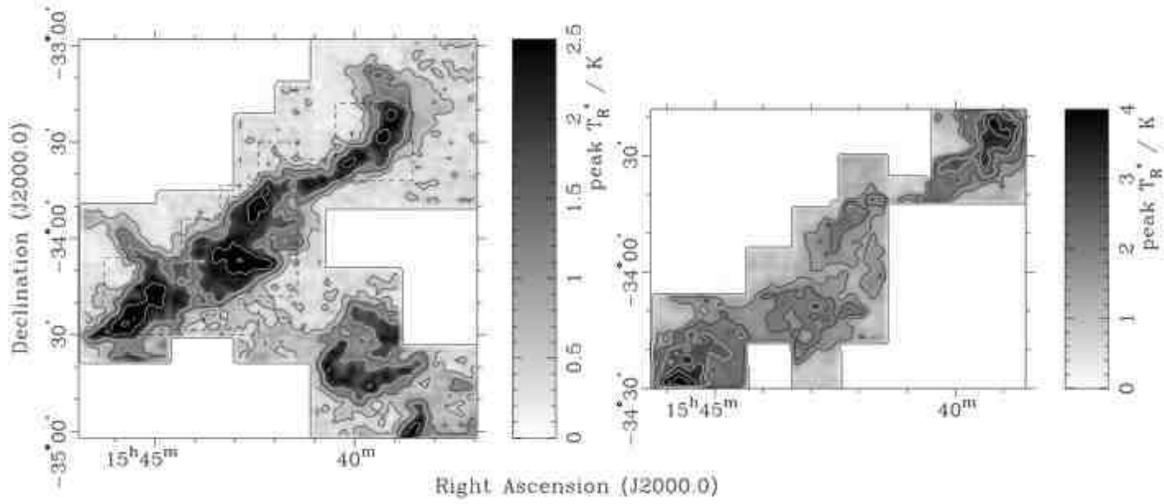}
\caption{Maps of the peak $T_R^*$ of (left) \thco\ 2--1 and (right) CO 4--3
toward Lupus~I. Left: contours at 0.5, 1\,K (black) and 
1.5, 2.5, 3.5\,K (white). Right: contours at 1.5, 2\,K (black) 
and 2.5, 3, 3.5, 4\,K (white). The broken outline denotes the extent
of the CO~4--3 map.}
\label{fig-lui-tmax}
\end{figure}

\begin{figure}
\includegraphics[scale=0.8]{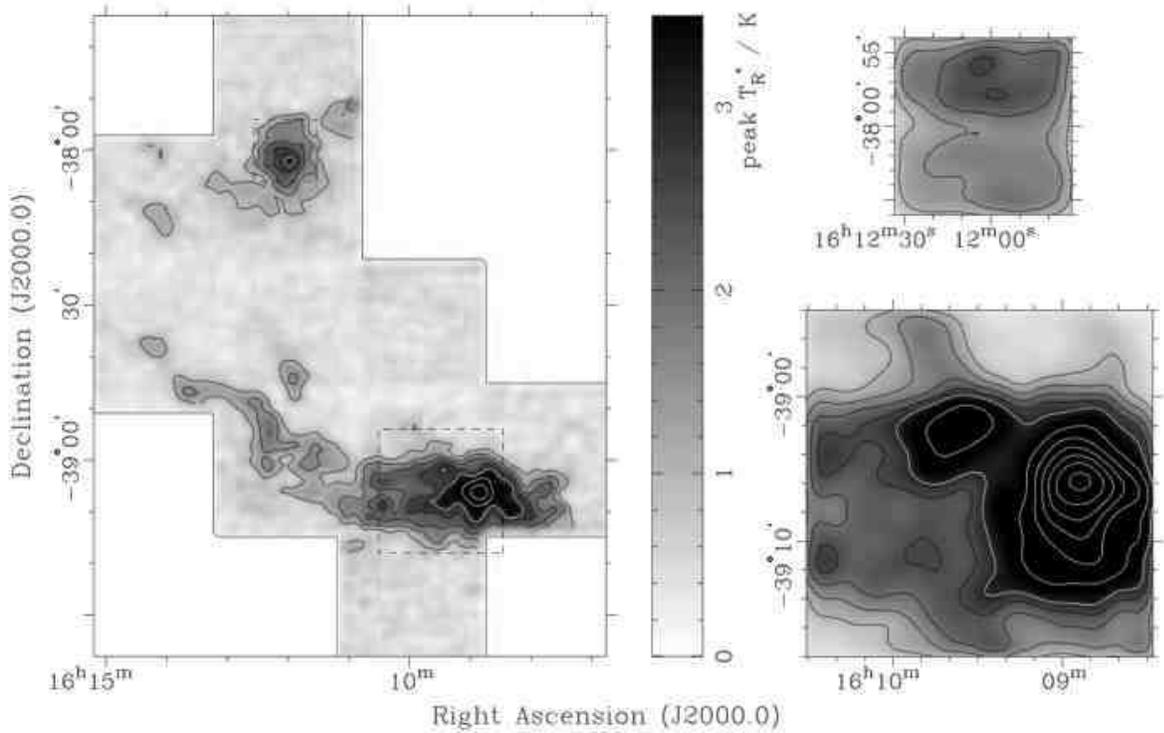}
\caption{Maps of the peak $T_R^*$ toward Lupus~III, as in 
Figure~\ref{fig-lui-tmax}: contours at 1, 1.5, 2, 2.5\,K (black) and 
3, 4,\dots 9\,K (white).}
\label{fig-luiii-tmax}
\end{figure}

\begin{figure}
\includegraphics[scale=0.8]{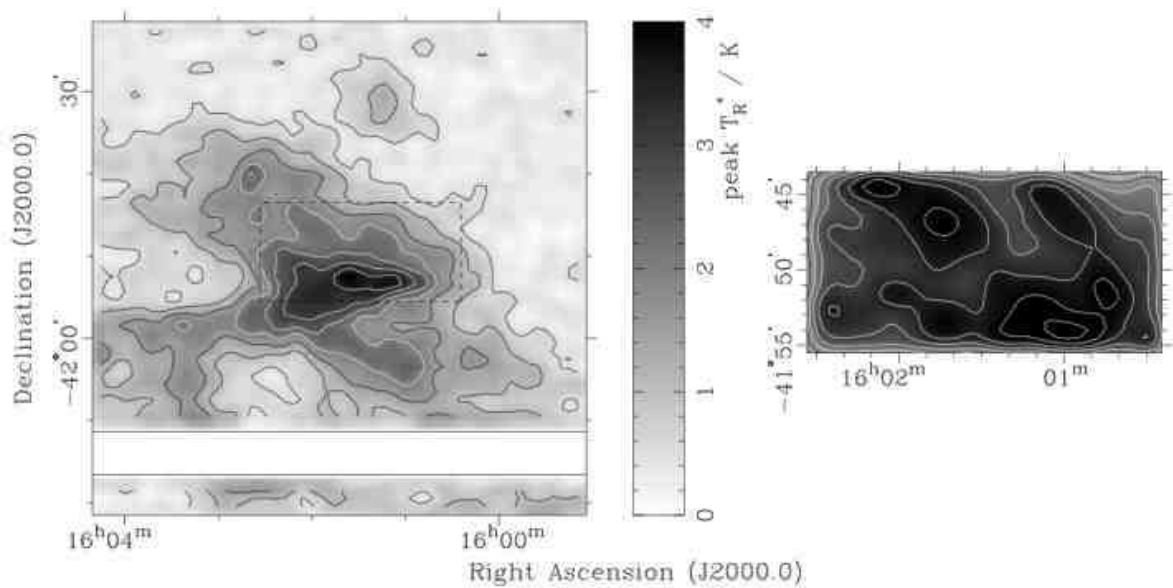}
\caption{Maps of the peak $T_R^*$ toward Lupus~IV, as in 
Figure~\ref{fig-lui-tmax}: contours at 0.5, 1, 1.5\,K (black) and 
2, 2.5,\dots 4.5\,K (white).}
\label{fig-luiv-tmax}
\end{figure}

Towards the B\,228 ridge, maps of extinction \citep{mips}, peak 
\trstar\ (Figure~\ref{fig-lui-tmax}), and integrated intensity 
(Figure~\ref{fig-lui-int}) are not well correlated. The extinction map is 
the most robust estimator of column density \citep{08goodman}: it shows 
a very strong peak in the middle of the ridge (DC 338.8+16.5), slightly 
less to the SE (339.2+16.1), and a broad low plateau to the NW 
(338.7+17.5). The \thco\ 2--1 peak \trstar\ map has similar peaks 
toward all three dark clouds, while the CO 4--3 peak \trstar\ map shows 
no significant peak toward 338.8+16.5 in the center. The integrated 
intensity map is very different from the extinction map, with the 
strongest emission in the NW followed by 339.2+16.1 in the SE, and 
338.8+16.5 comparatively weak in the center (hardly appearing in the 
CO 4--3 integrated intensity map).

\begin{figure}
\includegraphics[scale=0.64]{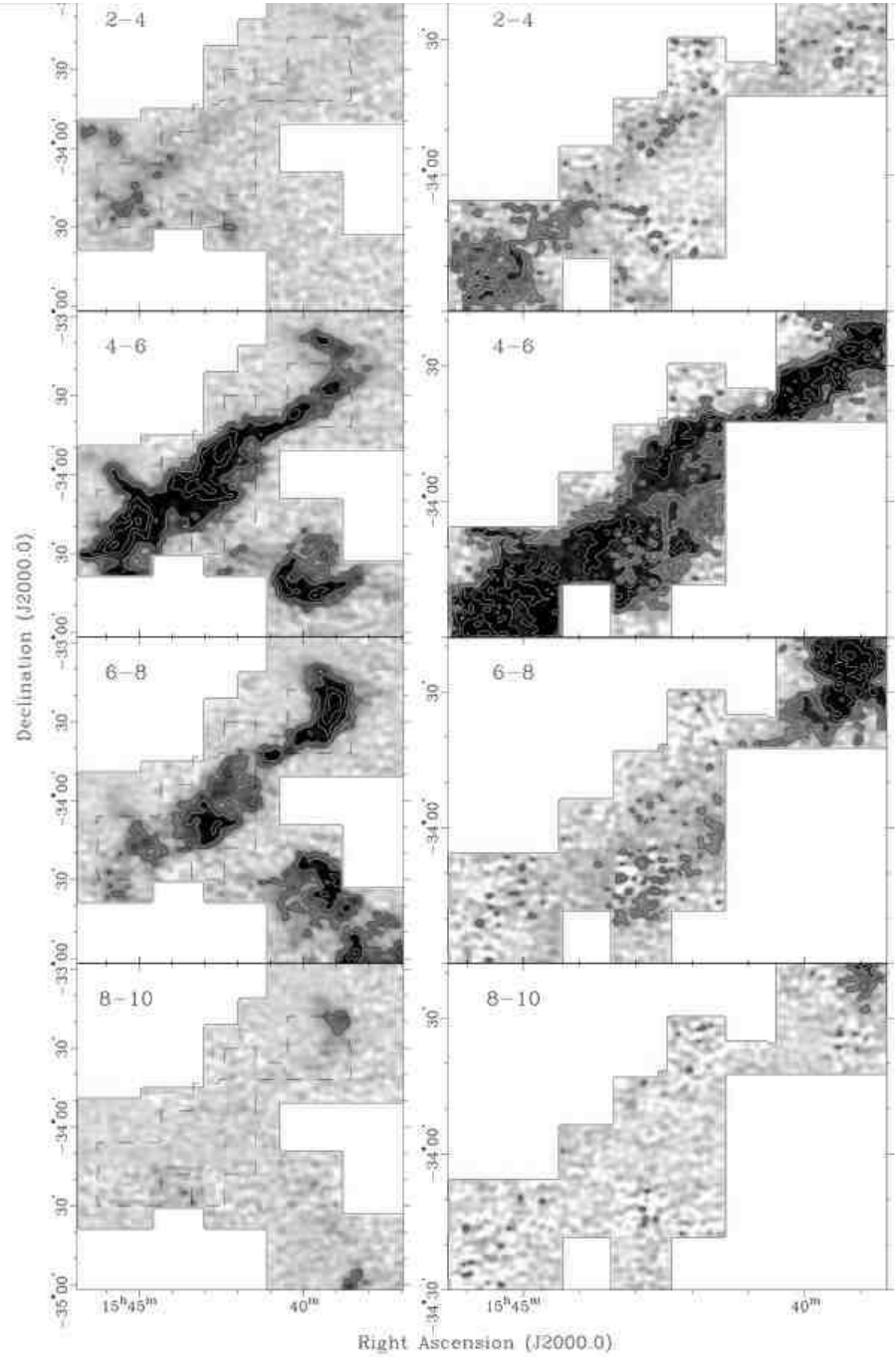}
\caption{Channel maps of Lupus~I in (left) \thco~2--1 and (right) CO~4--3,
labeled with $v_{LSR}$ range in \kms. Contours at 1\,\kkms\ (black) and 
2, 3,\dots 6\,\kkms\ (white). Broken outlines denote the extent of the 
CO~4--3 map.}
\label{fig-lui-chan}
\end{figure}

\begin{figure}
\includegraphics[scale=0.8]{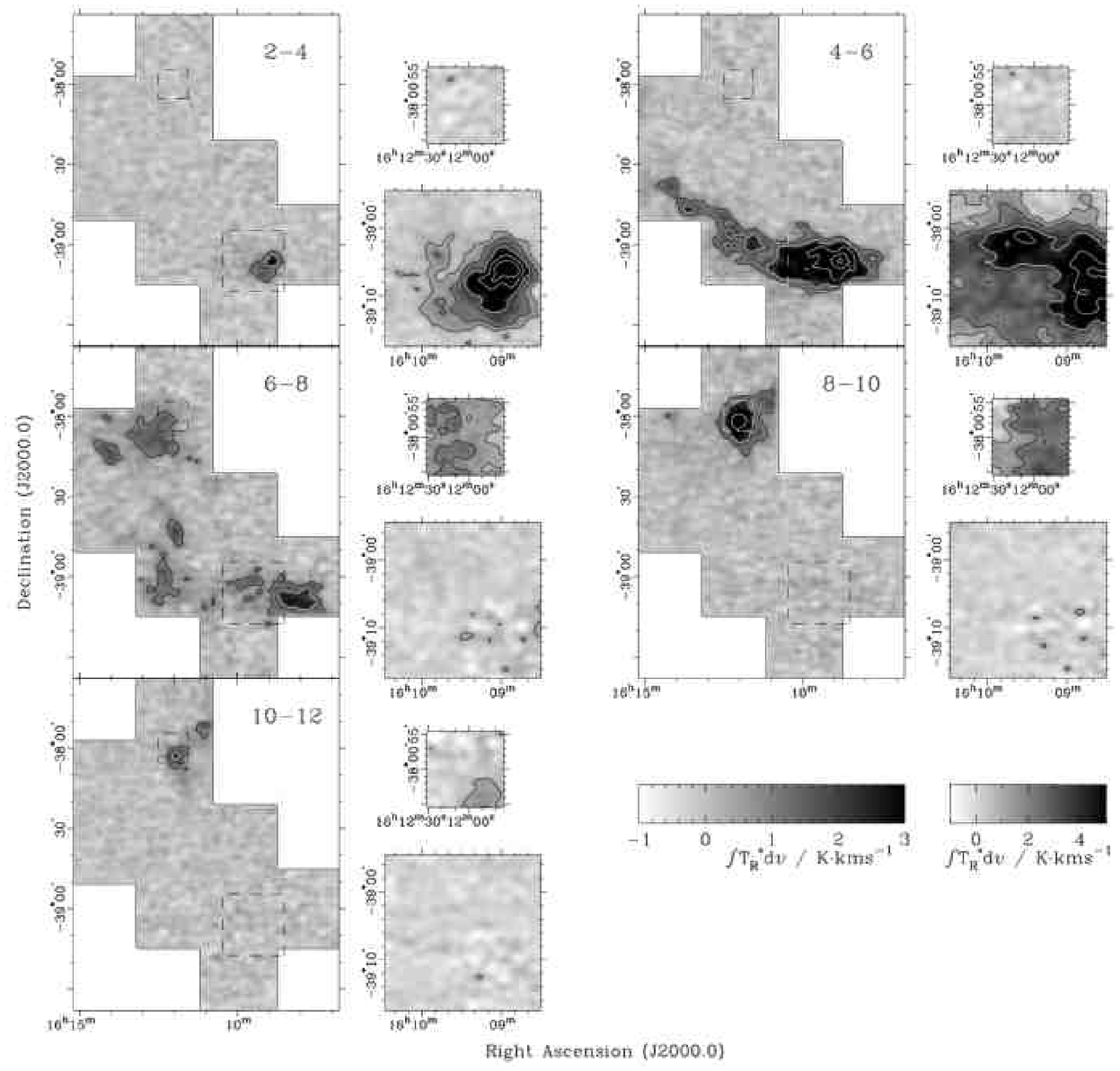}
\caption{Channel maps of Lupus~III, as in Figure~\ref{fig-lui-chan}.
\thco~2--1 contours are at 1\,\kkms\ (black) and 2, 4, 6, 8\,\kkms\ (white);
CO 4--3 contours are at 1, 2\,\kkms\ (black) and 4, 6, 8\,\kkms\ (white).}
\label{fig-luiii-chan}
\end{figure}

\begin{figure}
\includegraphics[scale=0.7]{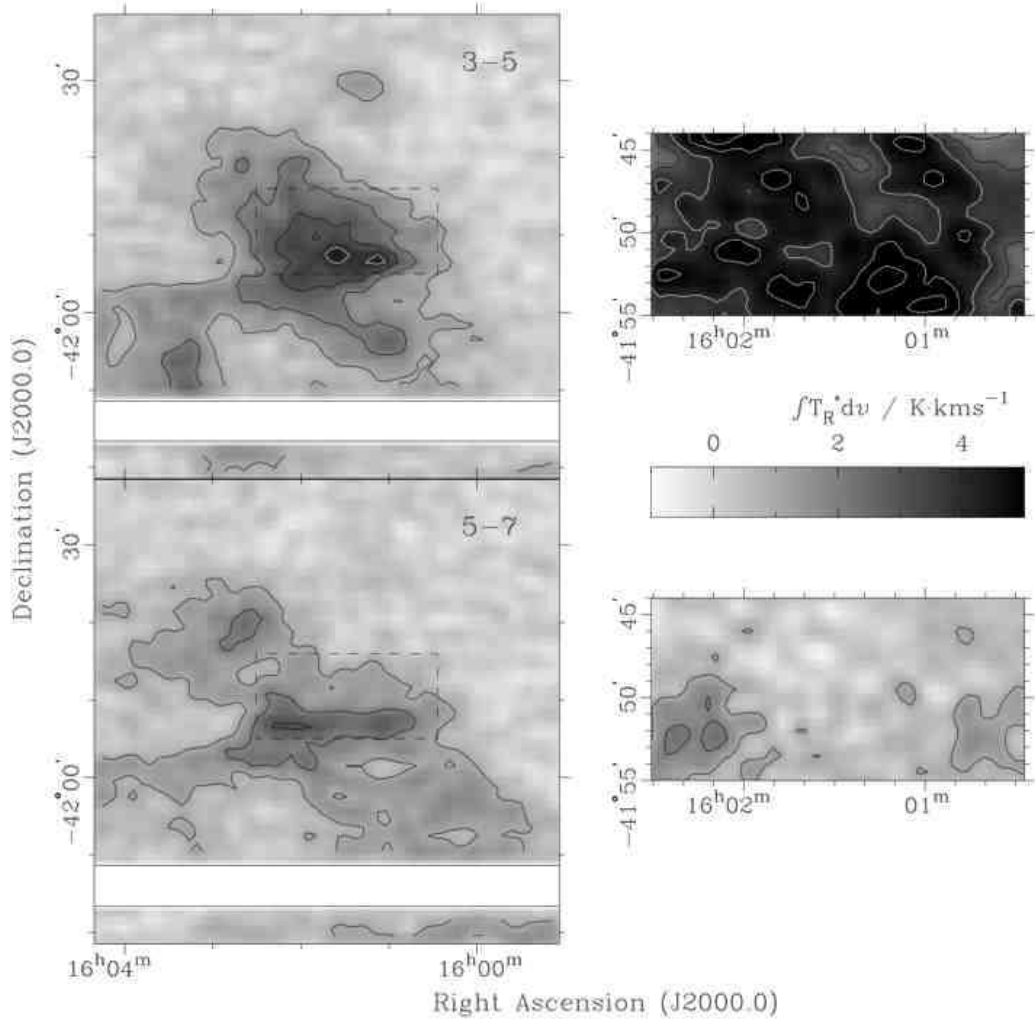}
\caption{Channel maps of Lupus~IV, as in Figure~\ref{fig-lui-chan}.
Contours at 1, 2, 3\,\kkms\ (black) and 4, 5, 6\,\kkms\ (white).}
\label{fig-luiv-chan}
\end{figure}

\begin{figure}
\includegraphics[scale=0.85]{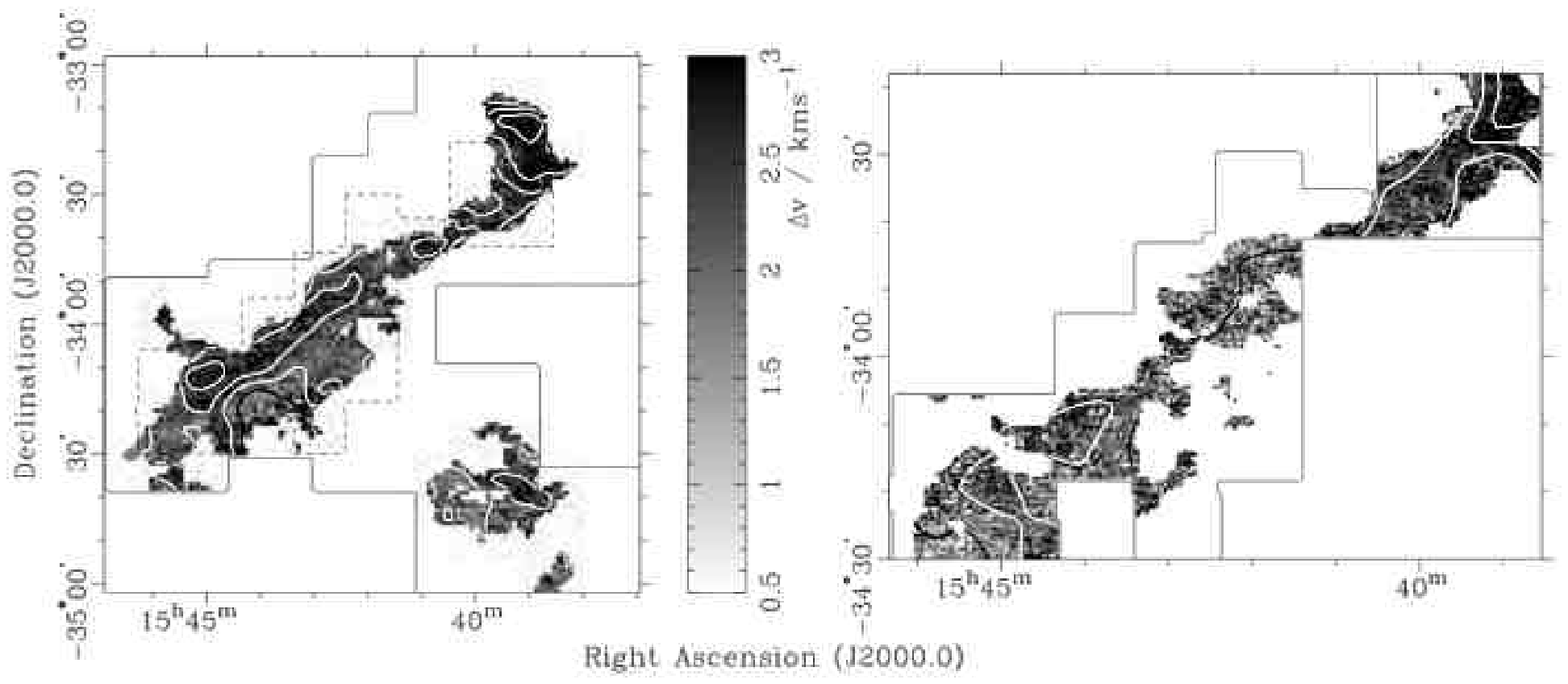}
\caption{Maps of the line widths of (left) \thco\ 2--1 and (right) CO 4--3 
toward Lupus~I. Contours are at 2, 2.4,\dots 3.6\,\kms. Gray scale is 
unsmoothed, while contours have been smoothed with a Gaussian kernel. 
Only spectra with S/N$>$10 are shown. Solid outlines indicate the extent 
of the maps, while the broken outline indicates the extent of the CO~4--3 map.}
\label{fig-lui-lwid}
\end{figure}

\begin{figure}
\includegraphics[scale=0.7]{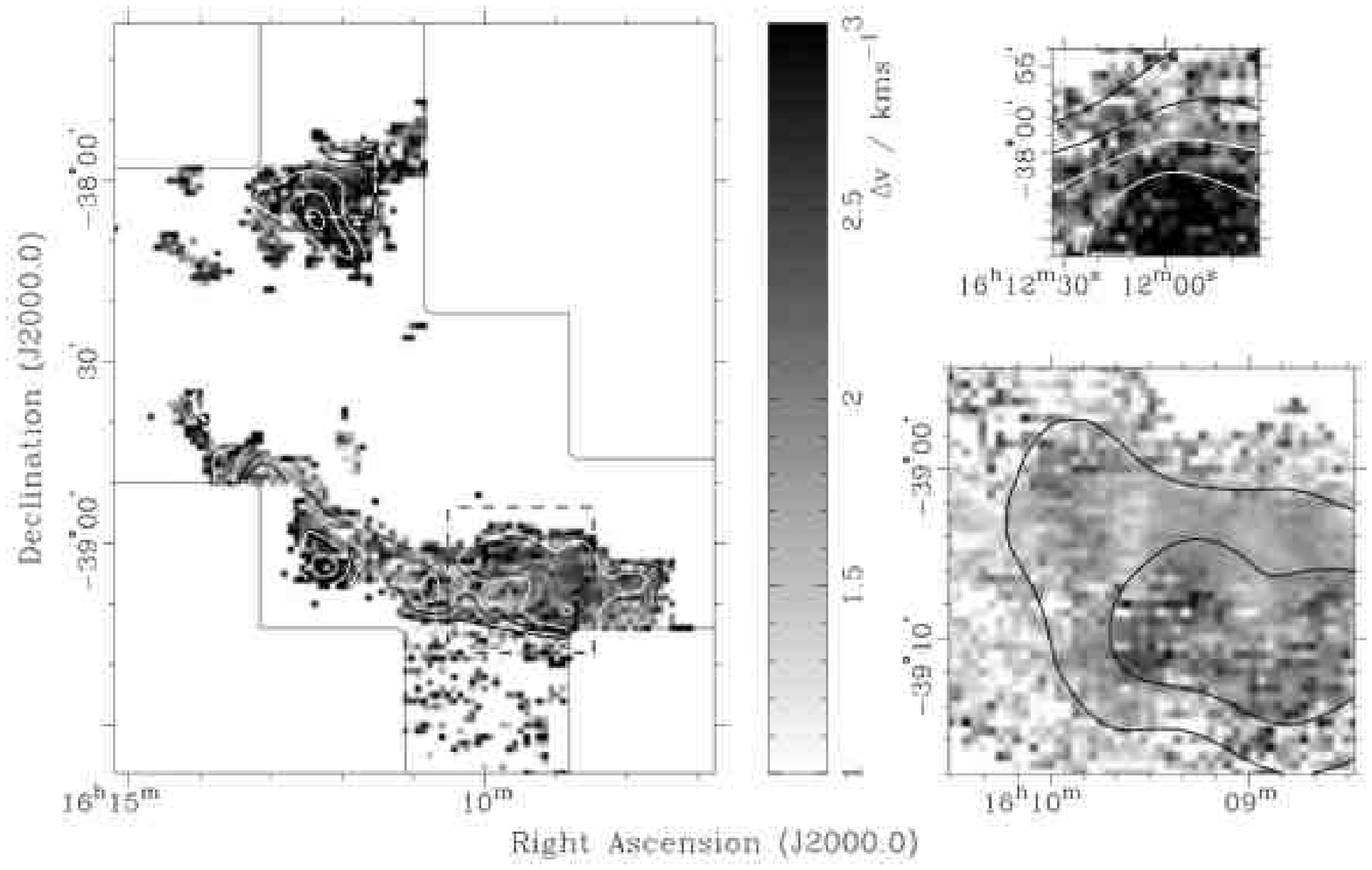}
\caption{Maps of emission line widths toward Lupus~III, as in 
Figure~\ref{fig-lui-lwid}: contours are at 1.5, 1.8, 2.1\,\kms\ (black) and 
2.4, 2.7, 3\,\kms\ (white). Only spectra with S/N$>$5 are shown.}
\label{fig-luiii-lwid}
\end{figure}

\begin{figure}
\includegraphics[scale=0.75]{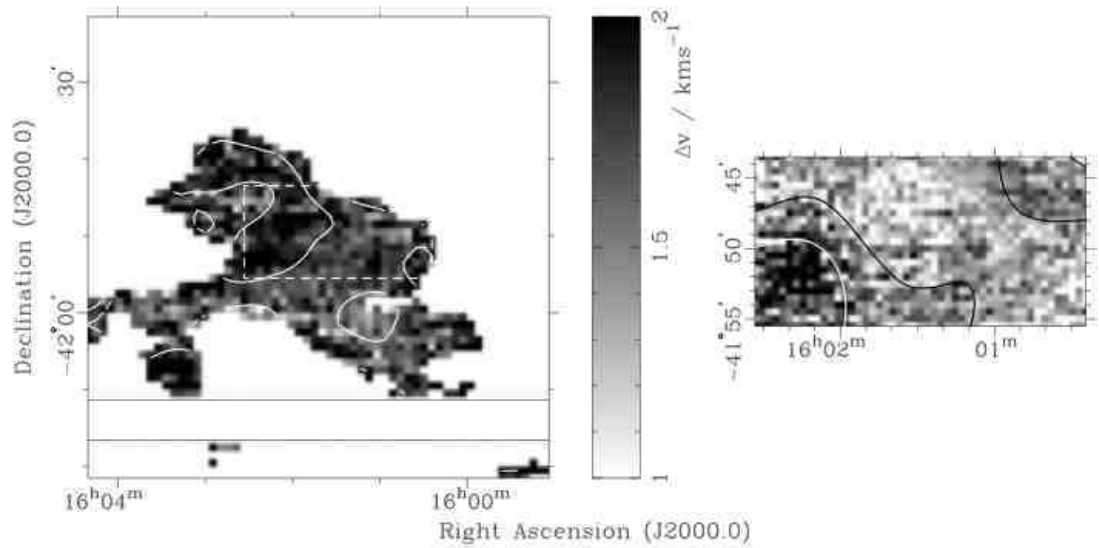}
\caption{Maps of emission line widths toward Lupus~IV, as in 
Figure~\ref{fig-lui-lwid}: contours at 1.4, 1.6\,\kms\ (black) and 
1.8, 2\,\kms\ (white). Only spectra with S/N$>$10 are shown.}
\label{fig-luiv-lwid}
\end{figure}

Previous molecular line studies have found a velocity gradient of about 
0.3~\kmspc\ along the B228 ridge \citep{01tachi,vmf}, but the channel and 
velocity maps (Figures~\ref{fig-lui-chan} and \ref{fig-lui-vcen}) show very 
little evidence of a smooth velocity gradient along the ridge. Rather, 
the two large clumps at the NW and SE end of the ridge have velocities 
that differ by about 3~\kms, while the ridge between them lies at an 
intermediate velocity. On the other hand, a velocity gradient across 
the center of the ridge shows up very clearly in both 2--1 and 4--3 
maps: the gradient of about 1~\kmspc\ is quite strong, compared not only 
to previously reported gradients in this cloud, but also to the gradients 
of 0.1--0.4~\kmspc\ found in areas of Taurus \citep{85murphy}. The 
gradient is smooth and coherent over the whole of this region --- 
an extent of more than 1~pc --- and appears quite linear, which would be 
consistent with solid-body rotation around the long axis of the filament 
\citep{alyssa}. A coherent pattern in this region may also be seen in 
the line width maps (Figure~\ref{fig-lui-lwid}), which show an increase in 
line width toward the leading (NE) edge of the ridge. There is no sign 
of a velocity gradient across the NW half of B228, which suggests that 
the NW part of the ridge may be distinct from the center.

One of the peaks in the center of the ridge, DC~3388+165-5, is 
associated with IRAS 15398-3359: this Class 0 YSO \citep{yancy} has a 
compact ($\sim$1\arcmin) molecular outflow with a dynamical timescale 
of about 2\,kyr \citep{96tachi}. It lies about 0.5~pc behind the steep 
edge of the ridge, in the E--W elongated core that forms the southern 
half of DC~338.8+16.5, and coincides with local maxima in integrated 
CO 4--3 emission (Figure~\ref{fig-lui-int}), in peak \trstar\ of both 
transitions (Fig~\ref{fig-lui-tmax}), and in CO 4--3 line width 
(Figure~\ref{fig-lui-lwid}). 

At the SE end of the ridge, the NE spur of DC~339.2+16.1 is blueshifted 
from the ridge, while the rest of the clump is at about the same 
velocity, suggesting that the spur might be a separate cloud superposed 
on the ridge. \citet{99lm} identified four optical extinction peaks in 
this clump: DC~3392+161-1 to -4. They associated Peak~1 with 
IRAS~15420--3408/HT~Lup, a CTTS \citep{comeron}, which is optically 
visible as a nebulous patch between peaks 1, 2 and 3 in 
Fig~\ref{fig-lui-opt}. The other 3 `starless' peaks were observed in 
CS 3--2 and DCO$^+$ 2--1 \citep{04lmp}: all three have 
\tastar(CS)$\sim 0.5$~K; DCO$^+$ was only detected (with 
\tastar(\dcop)$\sim 0.2$~K) toward peaks 3 and 4. DCO$^+$ is a 
high-density tracer, and its presence toward peaks 3 and 4 along the 
ridgeline, and absence from peak 2 at the start of the NE extension, 
suggest that, even in this complex structure, dense gas is concentrated 
to the ridge. The CO 4--3 data, with better spatial and velocity 
resolution, show additional structure in this clump: a patch of strong 
emission close to DC~3392+161-4, about 10\arcmin\ across, is 
blueshifted by about 1~\kms\ with respect to the main ridge structure, 
leaving a cavity in the emission at ambient velocity. Molecular outflow 
is a possible explanation, but since there is no obvious red lobe, this 
would require the YSO to lie behind the cloud. Blue asymmetries in 
molecular lines can be caused by infall, but \citet{04lmp} searched 
unsuccessfully for signs of infall toward DC~3392+161-4.

The parsec-scale ring-shaped structure to the SW of B\,228 is not covered by 
the CO 4--3 map. However, its \thco\ 2--1 emission looks quite similar to 
that seen in the bulk of the ridge, and the detection of DCO$^+$ toward the 
dense core embedded in the SE edge of the ring suggests a significant 
amount of dense gas. The structure may also have a strong velocity gradient: 
the channel maps (Figure~\ref{fig-lui-chan}) show two complementary 
semicircular structures in adjacent channels, i.e.\ separated by about 
2\,\kms, equivalent to a velocity gradient of at least 1~\kmspc. In the 
line width map (Figure~\ref{fig-lui-lwid}), the redder NW side of the structure 
has somewhat broader lines. \citet{04lmp} observed an extinction core at the 
SE edge of the ring: they found no signs of infall, but both CS 3--2 and 
\dcop\ 2--1 are stronger in this core than in any of the cores observed in 
B\,228. \citet{05teix} found a ring-shaped structure in extinction in 
Lupus~III, with a diameter of about 5\arcmin, which they interpret as the 
remnants of the molecular cloud that formed the nearby cluster of young 
stars. By contrast, this structure is larger (20\arcmin) and is not 
associated with a known cluster.

\subsection{Lupus III}

CO emission toward Lupus~III in both transitions is dominated by a 
compact source in the west of the cloud, close to core~F identified by 
\citet{05teix}. Proceeding from E to W are: Core~F; the compact CO 
source (slightly blueshifted from the bulk of the cloud); the 
HAe/Be stars HR\,5999 and 6000; and further \thco\ 2--1 emission 
extending E--W. This latter emission shows no particular sign of the 
ring structure found in extinction by \citet{05teix}, probably because 
the central void is smaller than the 220\,GHz beam. There is a 
significant velocity gradient across this area, of about 2~\kmspc, 
getting redder to the west, which is not simply due to the blueshifted 
emission of the compact source to the east.

To the east of the compact source, extinction cores A--E \citep{05teix}
blend together at this resolution: the integrated intensity map 
(Fig~\ref{fig-luiii-int}) shows two E--W spurs off the emission peak, the 
northern containing cores~C and D, and the southern containing E.
Further east still, the northern spur continues to core~B, while A lies just 
NE of DC~3397+92-3. The strongest \hthcop\ emission in this area 
($\sim 1$~\kkms) is found toward Core~E; Core~D, with about half the
\hthcop\ emission, contains the Class 0 YSO Lupus~3~MMS \citep{07tachi}.

Lupus~III\,N, the small ($<0.5$~pc) cloud core lying a few pc N of Lupus~III,
shows surprisingly complex structure: \thco\ 2--1 channel and velocity maps 
(Figures~\ref{fig-luiii-chan} and \ref{fig-luiii-vcen}) show the core to be 
lying in a larger ($\sim 2\times 1$~pc) structure, elongated SE--NW, with a 
significant velocity change over the long axis. The velocity gradient is
dominated by a sudden change just SE of the cloud core, also seen as a sharp 
edge in the reddest 2--1 channel map.

\subsection{Lupus IV}

The E--W elongated core of Lupus~IV has 3 peaks along its length: the 
central and western ones correspond to the extinction peak which 
reaches $A_V\sim 24$~mag. The peaks are most obvious in the \thco\ 
2--1 channel maps (Figure~\ref{fig-luiv-chan}); the peak \trstar\ map 
(Figure~\ref{fig-luiv-tmax}) shows them with much lower contrast than the 
extinction maps as the \thco\ transition is optically thick 
($A_V\sim 24$~mag corresponds to $\tau_0^{(13)}\sim$\,8--10). 
\citet{02my} found the \thco\ 1--0 line to be optically thick as well, 
saturating with respect to 100\,\micron\ IR emission. The eastern peak 
has broader lines yielding similar integrated intensity 
(Figure~\ref{fig-luiv-int}); at finer velocity resolution, these lines 
are found to be double peaked \citep{02my}. Although there is quite 
strong CO 4--3 emission throughout the core, and the velocity structure 
of the 4--3 emission is similar to that of \thco\ 2--1, the integrated 
4--3 emission is concentrated toward the E and W condensations only. 
Two elongated structures running NE of the core are also visible in the 
CO 4--3 peak \trstar\ map.

The strong velocity gradients seen in Lupus~I and III are absent from 
Lupus~IV. Some velocity structure is evident: the central condensation 
in the core and the two elongated NE structures are blueshifted (by 
1--2~\kms) compared to the E and W condensations. \citet{02my} reported 
significant velocity structure in the E--W direction, but our velocity 
maps show that this E--W structure is not found throughout the cloud.

\subsection{Comparison of CO transitions}

\subsubsection{Whole clouds}

\begin{figure}
\includegraphics{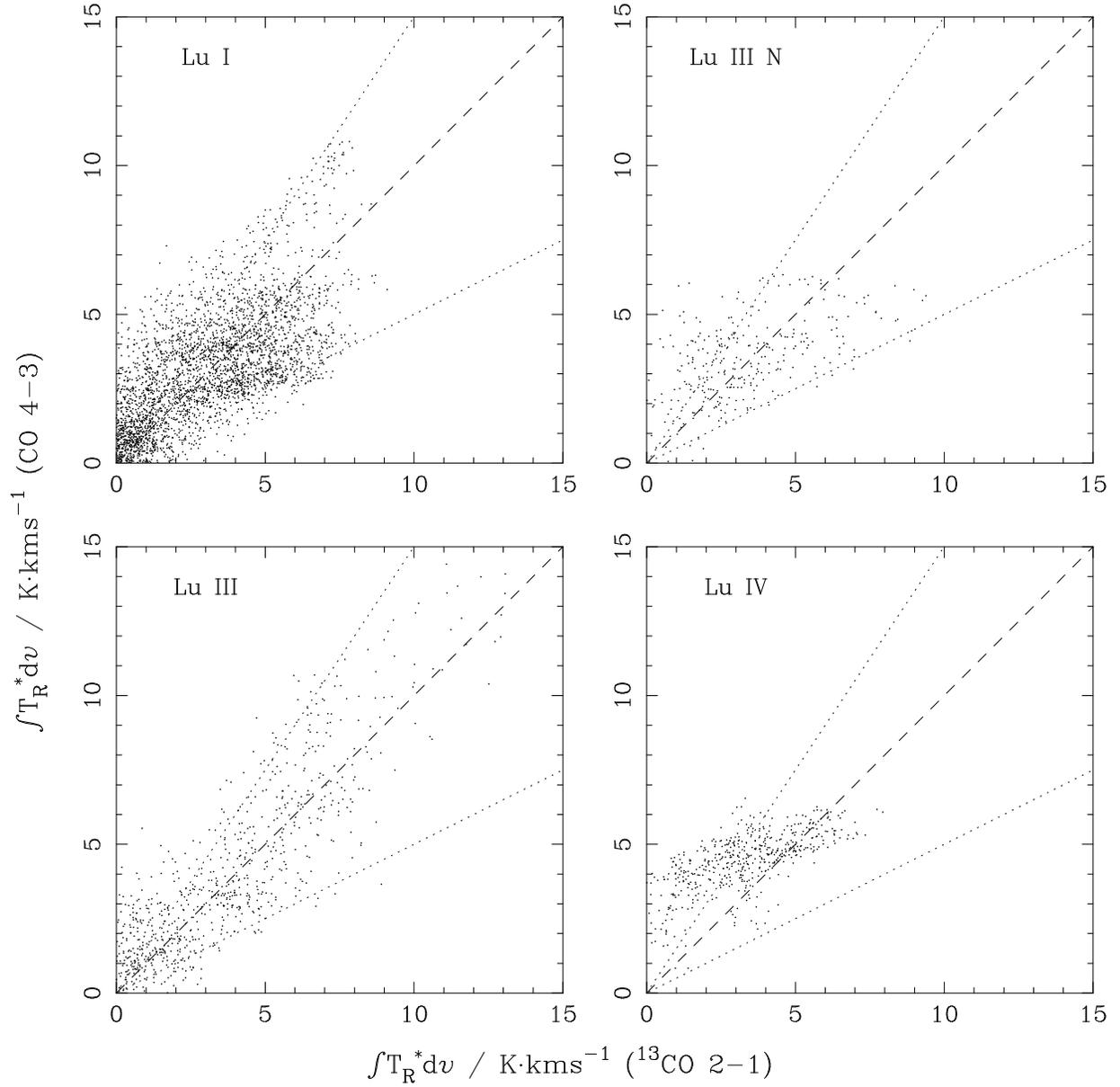}
\caption{Integrated intensities of \thco\ 2--1 and CO 4--3 for
all regions. The dashed line indicates a ratio of unity, while dotted
lines trace ratios (4--3/2--1) of 0.5 and 1.5.}
\label{fig-4321-int}
\end{figure}

\begin{figure}
\includegraphics{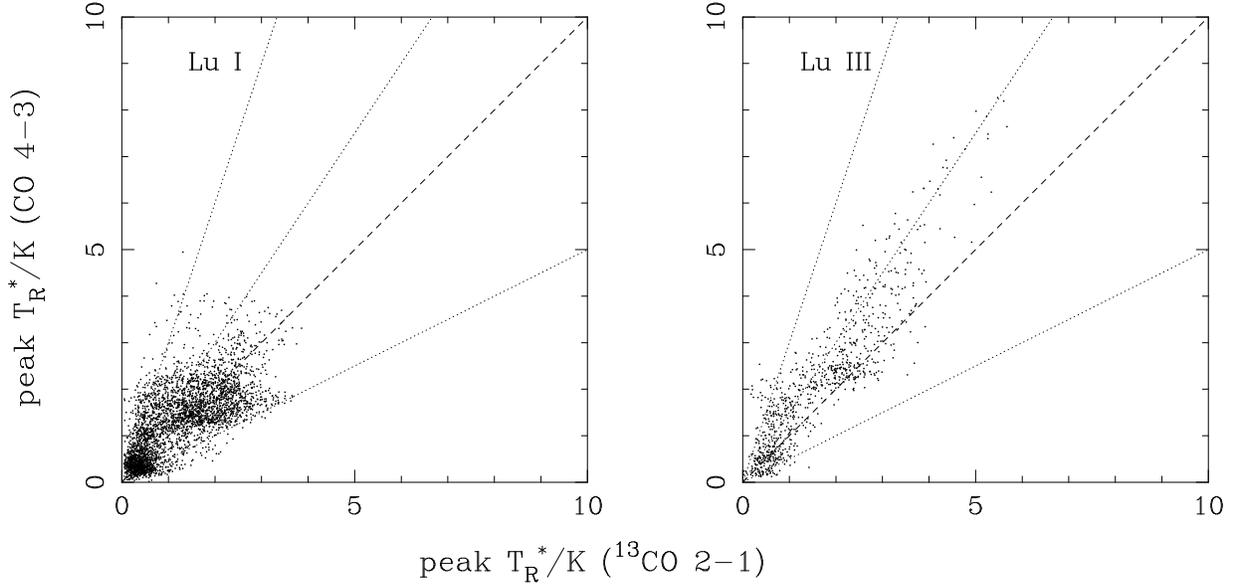}
\caption{Peak \trstar\ of \thco\ 2--1 and CO 4--3 for Lupus~I
and III only. The dashed line indicates a ratio of unity, while dotted
lines trace ratios (4--3/2--1) of 0.5, 1.5, and 3.}
\label{fig-4321-peak}
\end{figure}

CO 4--3 emission from the Lupus clouds can be compared to that of \thco\ 
2--1 by Gaussian smoothing the CO 4--3 map and plotting the integrated 
intensities (Figure~\ref{fig-4321-int}) and peak \trstar\ 
(Figure~\ref{fig-4321-peak}) against one another, point-by-point. The peak 
\trstar\ comparison is noisier than that of integrated intensity, and is 
only shown for Lupus~I and III, which have stronger emission.

The large 4--3 map of Lupus~I provides an overview of the bulk of the 
molecular gas. Clusters of points close to the origins of the integrated 
intensity and peak \trstar\ plots reflect the noise levels in the maps. 
The majority of the data are somewhat correlated, occupying regions of the 
plots bounded by (4--3)/(2--1)$\ga 0.5$ (both plots), 
$\int T_R^*{\rm(4-3)}dv\la 6$\,\kkms (integrated intensity), and 
$T_R^*{\rm(4-3)}\la 2.5$\,K (peak). The 4--3 emission is effectively 
saturated, while the \thco\ 2--1 emission takes a wide range of values. 
This suggests that even regions of quite low column density 
contain sufficient dense gas to emit optically-thick CO 4--3 radiation
(the critical density of CO 4--3 is of order $10^4$\,\percc, but see
Sec.~\ref{sec-disc-bulk}).

Another population of molecular gas can be identified toward Lupus~I: 
its \thco\ 2--1 peak \trstar\ is similar to that of the bulk population,
while its CO 4--3 peak \trstar\ is greater ($T_R^*{\rm(4-3)}\la 4$\,K). 
This suggests that the gas is warmer, but not denser than the bulk of the 
cloud. The integrated intensities of the transitions, however, are 
well correlated; this probably arises from increased line widths in both
transitions, and thus from increased velocity dispersion in the gas.
Inspection of the maps of integrated intensity, peak \trstar\ and line width 
toward Lupus~I (Figures~\ref{fig-lui-int}, \ref{fig-lui-tmax}, 
\ref{fig-lui-lwid}) shows that this component, with enhanced integrated 
intensity, peak CO 4--3 \trstar\ and line width, is found in the NW end of 
B\,228.

The much smaller CO 4--3 maps toward Lupus~III, III\,N, and IV provide
fewer data. However, two components can still be identified in Lupus~III: 
one component consistent with the Lupus~I bulk population, and the other 
with stronger emission. In contrast to its counterpart in Lupus~I, both 
integrated intensity and peak \trstar\ of the brighter component are 
well correlated. This brighter component constitutes the compact peak close 
to the HAe/Be stars in Lupus~III. Heating by the nearby stars could increase 
the peak \trstar\ and hence the integrated intensities of both transitions, 
without a significant change in the line width. The molecular gas in 
Lupus~III\,N and IV occupies a similar region of the integrated intensity 
plot to the bulk population of Lupus~I, suggesting broadly similar physical 
conditions.

Some positions show significant CO 4--3 emission without \thco\ 2--1: the 
spectra toward one of these positions are shown in 
Figure~\ref{fig-spectra-lui}, and the integrated intensity and peak \trstar\
plots show that this occurs in all the clouds. The combination of strong 
CO 4--3 emission and weak or absent \thco\ 2--1 implies warm gas with low 
column density; since the CO 4--3 map areas were chosen to cover \thco\ 
2--1 peaks, there may be more of this population outside our 4--3 maps.

\subsubsection{Dense Cores}

\begin{figure}
\includegraphics{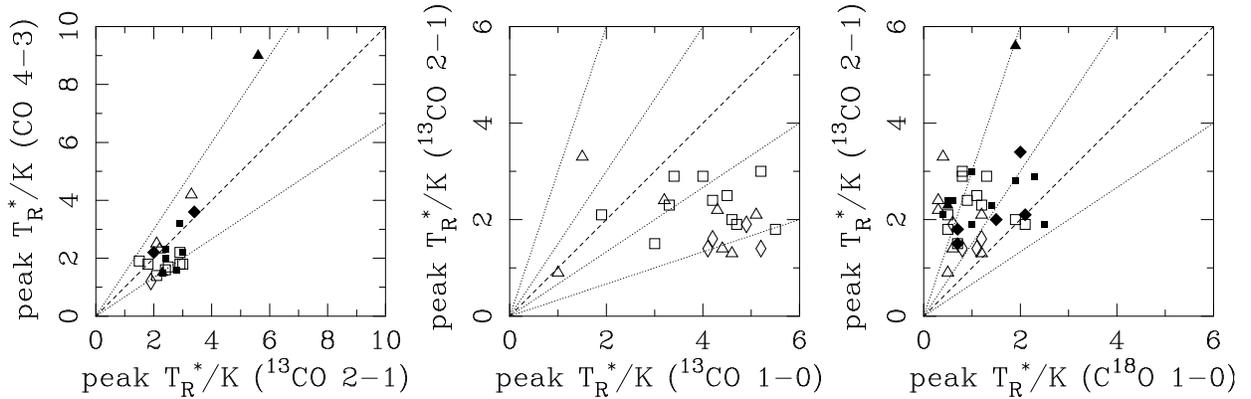}
\caption{Peak \trstar\ of CO transitions for cores seen in \ceo\ 1--0. The 
dashed line indicates a line ratio of unity, while dotted lines trace ratios 
of 0.67, 1.5 (left), 0.33, 0.67, 1.5, 3 (center), and 0.67, 1.5, 3 (right). 
Symbols denote cores in Lupus~I (squares), Lupus~III/III~N (triangles), and 
Lupus~IV (diamonds); filled symbols are \ceo -selected cores from 
\citet{99hara}, and open symbols are extinction-selected cores from 
\citet{vmf}.}
\label{fig-cotrans-peak}
\end{figure}

\begin{figure}
\includegraphics{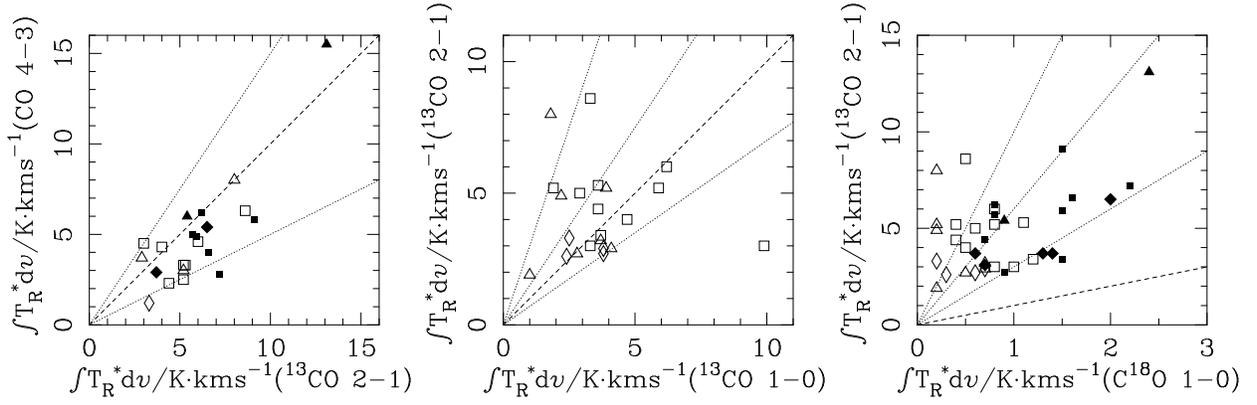}
\caption{Integrated intensities toward \ceo\ cores. The dashed line indicates 
a ratio of unity, while dotted lines trace ratios of 0.5 and 1.5 (left), 0.67, 
1.5, 3 (center), and 3, 6, 10 (right). Symbols are as in 
Figure~\ref{fig-cotrans-peak}.}
\label{fig-cotrans-int}
\end{figure}

Spectra toward specific cores have been measured by NANTEN and the 
Swedish-ESO Submillimeter Telescope (SEST): 
\citet{99hara} list the \ceo\ 1--0 line parameters for the spectrum at 
the peak of each dense core identified from their map, and \citet{vmf} 
list parameters for both \thco\ and \ceo\ transitions toward 
extinction-selected cores. Because the AST/RO maps are fully sampled, 
it is possible to obtain equivalent spectra (albeit with a beamwidth of 
3.3\arcmin, compared to the 2.8\arcmin\ NANTEN beam, and SEST's 
0.8\arcmin\ beam) and compare the peak \trstar\ 
(Figure~\ref{fig-cotrans-peak}) and integrated intensities 
(Figure~\ref{fig-cotrans-int}).

The peak \trstar\ in CO 4--3 and \thco\ 2--1 toward most of the cores 
lie in the same region of the plot as the bulk component of molecular 
gas described above. Four cores are clearly brighter than the rest: the 
two identified in Lupus~III are both associated with the bright compact 
source, the one in Lupus~I lies at the SE end of B\,228, and the one in 
Lupus~IV is in the middle of the central core. The majority of the 
extinction cores have \thco\ 2--1/1--0 peak \trstar\ ratios in the range 
0.3--0.7, inconsistent with the 0.7--2 expected from LTE 
(Appendix~\ref{app-co}). This could be caused by beam-dilution compared 
to the sub-arcminute SEST beam (but see below). The exception (one of 
the Lupus~III cores associated with the bright compact source) has a 
\trstar\ ratio of almost 3: it may be large enough not to suffer beam 
dilution, and warm enough to be close to the high-temperature line ratio 
limit. While the Lupus~I and III cores are well mixed in the plot, the 
Lupus~IV cores have rather low line ratios. There is no particular 
reason for them to be more beam diluted, so the lower line ratio may 
reflect a lower temperature. The plot of \thco\ 2--1 peak \trstar\ against 
that of \ceo\ 1--0 does not show any clear separation between the 
extinction and \ceo\ core samples. This is surprising, since the \ceo\ 
cores observed by NANTEN should suffer almost as much beam dilution as 
the AST/RO data, compared to the extinction-selected cores observed by 
SEST. The line ratios of the cores lie mainly between unity and $\sim 3$; 
correcting for the factor-of-2 \trstar\ discrepancy estimated above, this 
suggests line ratios of about 2--6, in line with LTE estimates for 
\ceo\ optical depths of 0.1--0.5 \citep{vmf}.

The CO 4--3 and \thco\ 2--1 integrated intensities of the cores are more 
scattered, but the high-\trstar\ cores in Lupus~III also have the highest 
integrated intensities. In contrast, the high-\trstar\ cores in Lupus~I 
and IV have more average integrated intensity, suggesting that the increased
\trstar\ (and hence temperature and/or density) is not accompanied by 
increased line width. The two more cores in Lupus~I with large \thco\ 2--1
integrated intensity also lie in the SE of B\,228. The ratios of the two 
\thco\ transitions (2--1/1--0, for the extinction-selected cores only) lie 
between about 0.7 and 1.5 (with a few between 1.5 and 3), which is 
consistent with our expectations from LTE, but not with the peak \trstar\
ratios above. Line parameters were estimated from the SEST data by fitting 
Gaussian line profiles \citep{vmf}, in contrast to the AST/RO and NANTEN 
data, for which integrated intensity and peak \trstar\ were measured 
directly, and the line width estimated by the ratio. This difference in 
analysis could cause a systematic discrepancy, with the SEST analysis 
yielding lower \trstar\ and broader line widths. The cores with higher ratios
(1.5--3) are all in Lupus~I or III, but are scattered throughout the clouds.
Ratios of \thco\ 2--1 to \ceo\ 1--0 integrated intensity are around 3--6
for the \ceo-selected cores observed with NANTEN (consistent with the 
peak \trstar\ ratios above), but the same ratios toward extinction cores
are generally higher. These high ratios generally arise from rather low
\ceo\ 1--0 integrated intensities ($\la0.5$~\kkms).

\subsection{CO Emission toward YSOs}

\begin{figure}
\includegraphics[scale=0.75]{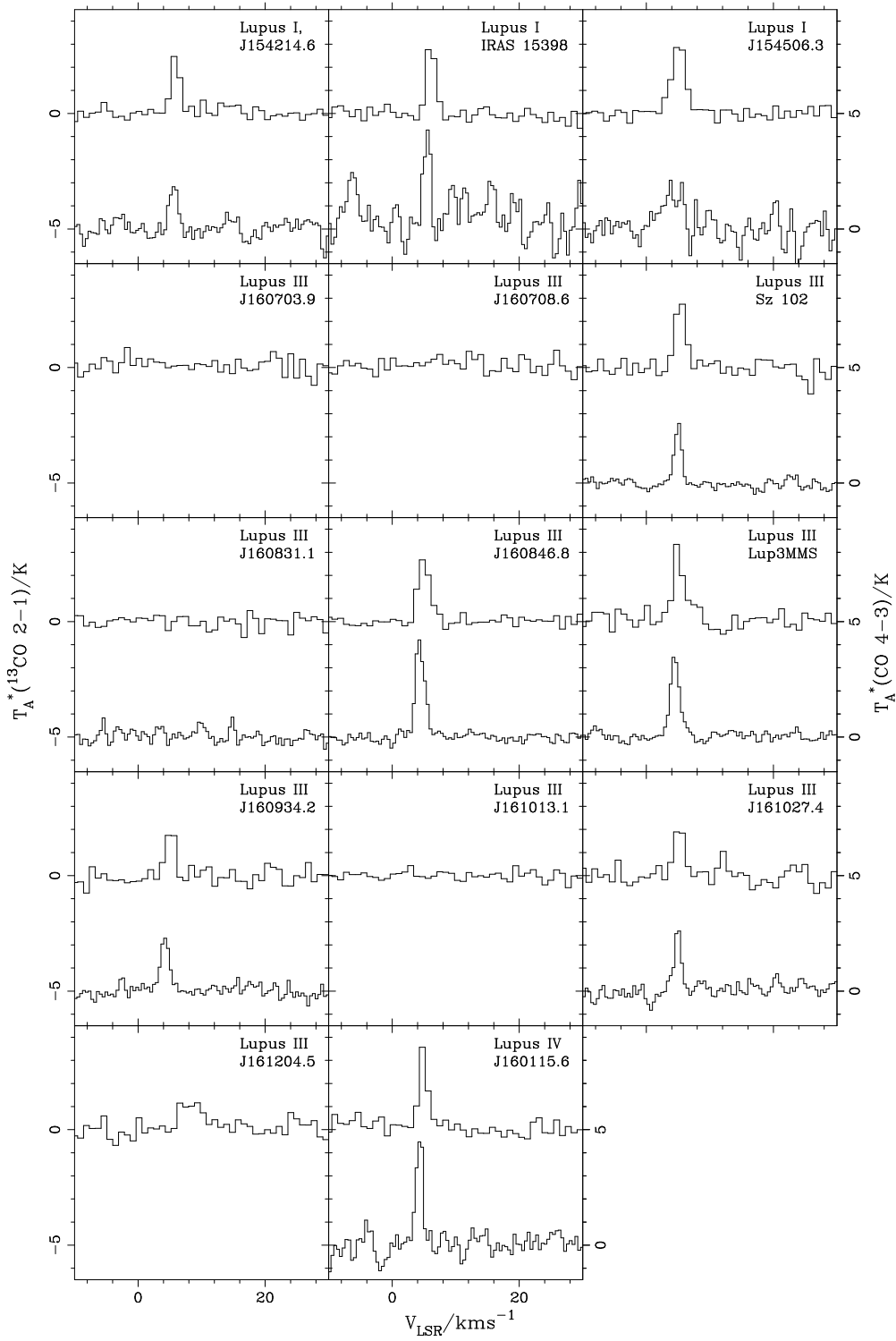}
\caption{Spectra toward the YSOs in Table~\ref{tab-yso}, as for 
Figures~\ref{fig-spectra-lui}--\ref{fig-spectra-luiv}.}
\label{fig-spectra-yso}
\end{figure}

The YSO population of Lupus is dominated by Class II and III objects 
\citep{comeron,merin}, but younger objects are also found there. The 
Class 0/I/F sources identified by \citeauthor{merin} lying within the 
\thco\ 2--1 maps are listed in Table~\ref{tab-yso}, together with the 
2--1 line parameters toward them and, where applicable, CO 4--3 
parameters; two objects that have been identified as background galaxies
\citep{09comeron} are excluded. Spectra toward the YSOs are shown
in Figure~\ref{fig-spectra-yso}: all 2--1 spectra are pointed within 
0.6\arcmin\ of the YSO position, and all 4--3 spectra are within 
0.25\arcmin. Spectra toward some YSOs do not show any significant 
emission: values of \trstar\ below $\sim1$K (in either transition) should 
be treated as noise.

Of the 4 YSOs in Lupus~III with very low or non-existent 2--1 emission,
\citet{09comeron} estimate 3 to have ages of order 100\,Myr. The rest of
the YSOs have line widths similar to those of the surrounding molecular
gas. This does not rule out the presence of line wings due to outflow: 
the Class 0 source Lupus~3~MMS has quite average line widths in both 
transitions, but the spectra themselves clearly show wings. The
flat-spectrum source J154506.3--341738 (in Lupus~I) has 
broader-than-average lines; nebulosity prevented it from being measured
in the optical by \citeauthor{comeron}.

\section{Discussion}

\subsection{The Bulk Molecular Material}
\label{sec-disc-bulk}

The estimates of column density, and hence mass, derived above assumed 
a blanket temperature of $9\pm1$\,K rather than the 13\,K assumed by 
\citet{96tachi} for Lupus~I and the 12--17\,K estimated toward 
Lupus~IV by \citet{02my}, both based on optically thick CO 1--0 emission. 
8\,K is the minimum excitation temperature 
consistent with the peak \thco\ 2--1 \trstar\ seen toward the 
majority of the gas in Lupus~I, III, and IV, and with the peak CO 4--3 
\trstar\ over most of the B\,228 ridge. Adopting excitation 
temperatures close to the minimum implies assumptions of optical 
thickness, LTE, and a filling factor close to unity, but is required 
for consistency with the extinction maps \citep{c2dproducts}: the 
$A_v=2.5$ and \trstar(\thco\ 2--1)$=1.5$\,K contours are quite similar. 
This implies $T_x\la 10$\,K for both PWE and FLW \thco-to-\av\ 
relations, although it does allow a higher $T_x$ toward Lupus~I for 
the Perseus \thco-to-\av\ parameters. 

The gas temperature estimate is consistent with 
estimates for many dark clouds. \citet{vmf}, using \thco\ and \ceo\ 
1--0 transitions, estimated the excitation temperatures of their sample of 
dense cores in Lupus to lie in the range 7--15\,K, with the majority of 
good estimates being colder than 10\,K; \citet{91clemens} estimated gas
temperatures toward a large sample of small dark clouds from CO 2--1, and
found that the majority were colder than 10\,K. Models of cold dark clouds 
yield dust temperatures $<$10\,K at the center \citep{01evans,01zucconi}. 
While the dust temperature is higher toward the cloud surface, gas 
temperatures are lower than dust temperatures for low column and volume 
densities \citep[e.g.,][]{97doty}, so the \thco\ 2--1 and CO 4--3 transitions
need not be dominated by the cloud centers. The CO 1--0 estimates mentioned
above are likely to be dominated by the cloud surfaces, which may be 
warmer than those of the majority of dark clouds due to external
heating by the nearby OB association. The peak CO 4--3 \trstar\ toward
Lupus~IV implies a $T_x$ of 10--12\,K, compared to $T_x\sim 8$\,K from
\thco\ 2--1. This may reflect a combination of external heating, as 
suggested by \citet{02my}, with high enough density to couple the gas and
dust temperatures more effectively than in Lupus I or III.

The column density of H$_2$ throughout most of the B\,228 ridge is within 
a factor of 2 of $5\times 10^{21}$~\percms \citep{mips}. The width of the 
ridge on the sky is about 10\arcmin, or 0.4~pc; if B\,228 is assumed to be 
a filamentary cloud, with a similar depth, the average volume density 
$\overline{n}$ is a few $10^3$~\percc. This is close to the critical density 
$n_c$ of the 2--1 transitions of CO and its isotopologues, supporting 
the assumption of LTE used throughout this work, but an order of magnitude 
lower than the critical density of CO 4--3 (a few $10^4$\,\percc, 
Appendix~\ref{app-co}). \citet{neal} found that significant emission in 
many transitions could arise from gas with volume density more than an 
order of magnitude lower than $n_c$ (although CO was not included in that 
study), but the similarity between the excitation temperatures of \thco\ 
2--1 and CO 4--3 suggests thermalized emission, and thus a volume density
close to $n_c$. The line emission is pervasive, which argues against its 
arising from cores or a high-density center of the ridge. However, if the 
bulk of the ridge material were taken to be close to the critical density 
of CO 4--3, the depth of the ridge along the line of sight would be an 
order of magnitude lower than its width in the plane of the sky, which 
seems implausible. It is more likely that the CO 4--3 emission arises from
a small fraction of dense gas found in clumps throughout the molecular cloud;
or from a thin warm shell around the outside of the cloud; or that subthermal
emission from the bulk of the molecular gas can account for the CO 4--3 
lines seen throughout the cloud. The latter possibility, in particular, may
be checked by modelling the emission.

\subsection{Cloud Cores}

\citet{vmf} estimated the optical depth of \ceo\ 1--0 toward their 
dark cloud sample to lie in the range 0.1--0.5 by comparing the \thco\ 
and \ceo\ lines, and assuming an abundance ratio of 5.5. A large 
fraction of their cores had line ratios in excess of 5.5, inconsistent 
with their assumed isotopologue ratio; however, ratios up to 8 are 
possible (Appendix~\ref{app-co}). \citet{99hara} estimated somewhat 
lower optical depths toward their \ceo\ cores, assuming $T_x=13$\,K; 
at $T_x=9$\,K, their estimated optical depth range becomes quite 
consistent with that of \citeauthor{vmf}. The \thco\ 1--0 optical depth 
toward the cloud cores is likely to range from 0.5 up to 2--4, depending 
on the abundance ratio.

\citet{05teix} estimated volume densities of a few $10^4$~\percc\ for 
the dense cores they identified in Lupus~III. If such densities are 
widespread among the dense cores in Lupus, the CO 4--3 transition is 
close to LTE, and it becomes possible to estimate the expected value 
of $(T_R^*)_{4-3}/(T_R^*)_{2-1}$ toward the cores. At $T_x=9$\,K, the 
high-column-density limit of the ratio (both transitions optically thick) 
is $\sim$0.4; as the optical depth of \thco\ becomes moderate, the ratio 
increases toward unity. However, at low optical depths, CO 4--3 may no 
longer be thermalized, making this estimate invalid. Most of the cloud 
cores have 4--3/2--1 ratios clustered around 0.5 
(Figure~\ref{fig-cotrans-peak}), with a few more at unity or above, and one 
over 1.5. It is difficult to achieve a line ratio significantly above 
unity under LTE: even at low optical depth, the temperature would need 
to be at least 20\,K.

\subsection{Lupus I}

The central condensation in the B\,228 ridge (338.8+16.5) has high 
extinction, moderate peak \trstar\ and integrated \trstar\ in \thco\ 2--1, 
and CO 4--3 emission quite similar to the bulk of the cloud. The line width 
is also quite similar to that of the bulk cloud, so the enhancement in 
\thco\ 2--1 is largely due to the increased peak \trstar. Elevated gas 
temperature would likely show up in the CO 4--3 emission, so the increased 
\trstar\ is largely due to the increased column density as the \thco\ 2--1 
transition becomes optically thick. This dense cloud seems to have more in 
common with the bulk material around it than it does with the emission peak 
to the NW, having similar temperature and line width. It is also associated 
with two embedded YSOs, including a known outflow source.

The strong, coherent velocity gradient seen north and east of 
338.8+16.5 runs perpendicular to the leading edge of the ridge (i.e., in
the direction of the \ion{H}{1} shell's expansion); the leading edge 
(i.e., toward the center of the shell) has a greater line width than the gas 
behind it. It is difficult to rule out 
temperature and optical-depth effects combining to mimic a velocity 
gradient, but similar patterns are seen in both \thco\ 2--1 and CO 
4--3 maps: a truly optically thin tracer is required to confirm the 
gradient. If the apparent gradient truly reflects the kinematics of 
the gas, the total change in velocity across the ridge is about 
1\,\kms, or half the line width. However, the change in velocity across 
a Jeans length (around 0.1~pc for these clouds) is only about 0.1\,\kms, 
which is unlikely to add significant support against collapse. Indeed, 
338.8+16.5 seems to be part of this velocity field, and contains a known 
Class~0 YSO.

The integrated intensity maximum at the NW end of B\,228 is due to 
the combination of enhanced peak \trstar\ and broader lines. Throughout 
this area, the peak CO 4--3 \trstar\ is $\ga 3$\,K, implying a gas 
temperature of at least 10\,K. Although the 4--3 emission probably does 
not sample the whole of the gas column, there is supporting evidence for 
a warmer temperature: the {\em Spitzer} IR maps \citep{mips} show this end of 
the filament to be bluer than the central part, with strong 24\,\micron\ 
emission \citep[see also][]{merin}, which could be due to a higher dust 
temperature. No Class~0/I/F YSOs are found in the northern part of the 
ridge, and only one Class~III \citep{mips}, so there are no obvious
internal heating sources. The nearby Upper-Sco OB subgroup lies to the
NE of Lupus~I, but is unlikely to be causing the heating, since no
similar effect is to be found in the center of the ridge.

At the SE end, the picture is complex: the extinction and peak \trstar\ for
both transitions are high at the SE end of 339.2+16.1, while line width and 
integrated intensity peak further NW. The strong enhancement of peak CO 4--3
\trstar\ at the SE end suggests increased temperature as well as column
density. The associated YSOs all lie to the NW in this area, so internal 
heating seems unlikely.

\subsubsection{Interaction with the Upper-Sco Shell}

The \ion{H}{1} emission from the Upper-Sco shell lies at a similar 
velocity \citep[3--9\,\kms,][]{92degeus} to the Lupus clouds, consistent 
with their being in contact with one another, and Lupus~I being dynamically 
affected by the shell. If such an interaction is going on, 
the combination of continuity and conservation of momentum require that
the sum of pressure and momentum flux is conserved by the interaction.
The limitations of both molecular and atomic data make it impossible to 
look for clear diagnostics of interaction between the shell and ridge,
but some rough estimates can be made. For both phases,
$$\frac{\left(P+\rho v^2\right)/k}{\rm cm^{-3}K} = 
\mu\left(\frac{n}{{\rm cm}^{-3}}\right)
\left(\left(\frac{\sigma_v}{{\rm km\ s^{-1}}}\right)^2+
\left(\frac{v}{{\rm km\ s^{-1}}}\right)^2\right)$$
where $\mu=170$ for \ion{H}{1}, and 340 for \htwo. Based on the \ion{H}{1}
data \citep{92degeus}, this quantity may be estimated to be of order 
$10^6$\,cm$^{-3}$K, about 90\% of which is the momentum flux component
due to the expansion velocity of the shell \citep[10\,\kms,][]{92degeus}.
$P+\rho v$ for the molecular gas in Lupus~I is
also around $10^6$\,cm$^{-3}$K, but is approximately evenly divided between
pressure and momentum flux (due to velocity gradients).

The rough similarity in this sum between the \ion{H}{1} and molecular gas
is consistent with interaction between them. If the \ion{H}{1} shell is 
indeed affecting Lupus~I, it is likely doing so through the momentum flux
of its expansion. This transfer of momentum could be causing the velocity
gradient across the B\,228 ridge (in the direction of the \ion{H}{1} 
shell's expansion); this process could be analogous to the `streamers'
in the Ophiuchus complex \citep{92degeus}. The enhanced line width at the 
NW end of B\,228 and along the leading edge of the ridge (implying increased
pressure in the molecular gas) is also consistent with the effect of
the \ion{H}{1} shell. More detailed studies of both \ion{H}{1} and
molecular gas may support the idea of Lupus~I being affected by the 
Upper-Sco shell: better estimates of the volume density of \htwo\ and
higher-resolution maps of \ion{H}{1} are required.

\subsection{Lupus~III}

The brightest part of Lupus~III near HR\,5999 and 6000 was mapped by
\citet{07tachi} in the millimeter-wave continuum, \ceo\ 1--0 and \hthcop\ 1--0, 
all with higher resolution than the AST/RO data. The \ceo\ emission peak 
coincides with the CO peak in our maps, and extends to the east, with 
another E--W elongation a few arcminutes to the north. The continuum map, 
together with a near-IR extinction map \citep{00nakajima}, shows the same 
southern E--W structure extending further east, while the northern 
structure breaks up into two clumps, the denser western one containing a 
Class~0 protostar (Lupus~3~MMS). In the \hthcop\ map, there is no 
emission at the CO peak; the emission peaks strongly to the east in the 
southern E--W structure, while the clumps in the northern structure show 
up as smaller peaks. 

\citet{07tachi} suggest that the lack of \hthcop\ emission at the CO peak
can be accounted for by a long path length through gas with volume density
significantly lower than the critical density ($\sim 10^5$\,\percc\ for 
\hthcop). They estimate a column density of $1.2\times 10^{22}$\,\percms,
equivalent to \av\ of about 13~mag, which is consistent with the extinction
maps \citep{mips}. The peak \trstar\ in CO 4--3 (about 9\,K) implies
$T_x\ga 18$~K, while the 6\,K peak \trstar\ of \thco\ 2--1 only requires 
$T_x\ga 11$~K. Compared to the rest of the Lupus clouds, where the CO 4--3 
and \thco\ 2--1 excitation temperatures are quite similar, this is a 
significant discrepancy. It could be explained by the CO peak being compact, 
as suggested by the CO 4--3 peak \trstar\ map, so that the \thco\ 2--1 
measurement is beam-diluted. Alternatively, the CO 4--3 could be tracing an 
outer shell heated by the HAe/Be stars. However, the \thco\ 2--1 emission is 
also optically thick (for $T_x=11$\,K, \av\ of 13~mag implies an optical 
depth of about 8), and so will trace similar material. These temperature 
estimates tend to support the argument that the gas is too warm for 
depletion of \hthcop\ to explain the lack of emission toward the CO peak 
\citep{07tachi}, although the estimates are unlikely to apply to the center 
of the clump. The strong CO 4--3 emission implies that the transition is 
thermalized, so a significant amount of the gas in the clump must have 
volume density $\ga 3\times 10^4$\,\percc. While \citet{07tachi} derived an 
average volume density of $10^4$\,\percc, there must be enough significantly 
denser gas to thermalize the CO 4--3 line, but not enough to excite the 
\hthcop\ 1--0 transition.

In projection, Lupus~III lies far away from any part of Sco-Cen 
\citep{01tachi}, in contrast to Lupus~I and Lupus~IV (discussed below), so 
the HAe/Be stars probably influence it far more than the 
OB associations. However, there is evidence for Lupus~III being further 
away than Lupus~I and IV, possibly as much as 50~pc \citep[e.g.][]{comeron}. 
If this is the case, then Lupus~III could lie behind Sco-Cen, and could be 
affected by either or both of Upper-Sco and Upper-Cen-Lup. The large 
line widths seen toward Lupus~III\,N could be caused by external influence
in the same way as the broader lines seen toward Lupus~I.

\subsection{Lupus IV}

The peak \trstar\ in CO 4--3 toward Lupus~IV of about 4\,K, implying 
$T_x\ga 12$~K, occurs around the outside of the extinction peaks, which
have peak \trstar\ around 3\,K ($T_x\ga 10$\,K). This suggests that the
outside of the clump is significantly externally heated. The \thco\ 2--1
peak \trstar\ implies $T_x\ga 8$\,K, but even this measure is unlikely to
sample the extinction maxima properly, since it will be dominated by the
outer layers of the structure. The CO 4--3 temperature estimates are in 
line with those seen at the NW end of B\,228, where Lupus~I seems to be
strongly affected by Upper-Sco. Lupus~IV is on the opposite side of the 
Lupus complex to Lupus~I, Upper-Sco and its \ion{H}{1} shell 
\citep{01tachi}, but faces the Upper-Cen-Lup subgroup, which lies to the 
W and SW. \citet{02my} suggested that Lupus~IV was shaped by the 
influences of both subgroups, and noted that some of the velocity 
gradients they saw in Lupus~IV were along the vector toward 
Upper-Cen-Lup. Much of the enhancement in the peak CO 4--3 \trstar\ lies 
on the S and W sides of the extinction peaks, which is consistent with a 
picture of external heating by the radiation field from Upper-Cen-Lup. 
There is, however, no nearby \ion{H}{1} shell, the Upper-Cen-Lup shell 
having passed the Lupus clouds long ago \citep{02my}. If interaction with 
the \ion{H}{1} shell causes the enhanced line width at the NW end of 
B\,228, the lack of any such interaction in Lupus~IV would be consistent 
with its rather low line widths.

\section{Conclusions}

Fully sampled degree-scale maps of the \thco\ 2--1 emission toward 
the Lupus I, III, and IV clouds trace the column density and temperature 
of the gas, the transition becoming optically thick in the cloud cores. 
The peak \trstar\ is well correlated with the near-IR extinction 
\citep{mips,c2dproducts}, and a comparison of the two suggests that the
bulk of the molecular gas in Lupus has a temperature of 8--10\,K, 
rather than the 10--17\,K generally adopted elsewhere 
\citep[e.g.][]{96tachi,02my,05teix}. This estimate is fairly robust to 
changes in the relationship between \thco\ column density and \av. 
Estimates of the cloud masses from the \thco\ maps are reasonably 
consistent with those derived from extinction mapping. The differences 
between these estimates vary greatly from cloud to cloud, and suggest that 
there may be significant spatial variation in the \thco-to-\av\ 
relationship, as found in the Perseus complex \citep{08goodman}. The 
line widths of \thco\ 2--1 toward the clouds are higher than previous 
estimates \citep{c2dresults}, around 2\,\kms, with Lupus~III\,N rather 
broad and Lupus~IV rather narrow.

Fully sampled CO 4--3 maps covering most of Lupus~I and small regions of
Lupus~III and IV trace dense gas: the peak \trstar\ generally indicates 
excitation temperatures quite close to those of \thco\ 2--1, and hence 
that the transition is largely thermalized. This suggests that the volume 
density $\ga 10^4$\,\percc, although modelling will be required to ascertain 
the required density. CO 4--3 emission is pervasive toward Lupus~I (the map
of which covers a large area), implying that this dense gas is found either 
throughout or all around the outside of the cloud, although it may 
comprise a fairly small fraction of the cloud mass.

The physical conditions of the molecular gas vary along the B\,228 ridge in 
Lupus~I. At the NW end, the gas has broader lines and probably higher 
temperature than in the bulk of the cloud; the column density
is not particularly high and there is only one Class~III YSO. In the center 
of the ridge, the dark cloud 338.8+16.5 is associated with recent star 
formation \citep{96tachi,yancy}; in this area a coherent velocity gradient 
of about 1\,\kmspc\ runs across the ridge. The SE end of the ridge is 
complex, with YSOs, enhanced line width and integrated intensity on the NW 
side, and column density (and possibly temperature) peaking to the SE.
The enhanced line widths and velocity gradient in B\,228 are consistent
with a dynamical interaction between Lupus~I and the \ion{H}{1} shell 
around the Upper-Sco subgroup of Sco-Cen \citep{92degeus,01tachi}.

To the north of Lupus~III, the small cloud Lupus~III\,N has similar 
characteristics to the bulk of the other clouds, albeit with broader 
lines and significant velocity structure. Lupus~III itself contains a 
compact CO peak which is probably heated by the nearby HAe/Be stars 
HR\,5999 and 6000. The gas in this clump contains sufficiently dense gas 
to thermalize the CO 4--3 transition ($n_c\sim 3\times 10^4$\,\percc), 
but not to thermalize the \hthcop\ transition mapped by \citet{07tachi}, 
which would require $n\ga 10^5$\,\percc. The rest of Lupus~III seems to 
have quite similar physical conditions to those in the rest of the 
clouds, and shows no particular sign of being affected by the nearby 
OB subgroups.

Lupus~IV contains peaks of very high column density \citep{mips} associated
with slightly warmer gas temperatures (10--12\,K). These temperatures
are estimated from the optically-thick CO 4--3 transition, which is strongest
around the extinction cores, suggesting significant external heating.
The average line width of \thco\ 2--1, however, is significantly lower 
than those of Lupus~I and III. Lupus~IV faces the Upper-Cen-Lup subgroup of 
Sco-Cen and \citet{02my} suggested that it is influenced by the OB stars; 
the Upper-Cen-Lup \ion{H}{1} shell passed by the Lupus clouds a few Myr 
ago, so Lupus~IV is more likely to be affected by the radiation field from 
the OB stars.

Despite the basic similarities in their physical conditions, the three
clouds have significant differences: Lupus~I appears to be strongly affected
by external thermal and dynamical influences from the nearby 
Upper-Sco OB association, and does
not display widespread star formation. Lupus~III shows no sign of external
influence --- parts of the cloud are heated internally by its own young
stars. Lupus~III\,N seems entirely quiescent, yet has a large average
line width. Lupus~IV has the greatest column density and the narrowest
average line width, has almost no star formation as yet, and may be
heated externally by the Upper-Cen-Lup OB association.

A detailed spatial comparison of CO and extinction maps will yield more 
accurate estimates of the physical conditions of the Lupus clouds, as well
as mapping the variation in the \thco-to-\av\ ratio. Mapping
of additional CO transitions is crucial to the understanding of the clouds:
more optically-thin lines (\ceo\ and even $\mathrm{C^{17}O}$) are 
particularly important. Maps of the Upper-Sco \ion{H}{1} shell with 
comparable resolution to the Lupus maps are required to look for more
definitive signs of interaction between the shell and the molecular clouds.
More sophisticated radiative transfer calculations
are beyond the scope of this work, but are required to use the CO 4--3 
emission as a proper constraint on the physical conditions of the gas,
and hence to understand the structure of the Lupus clouds and how they are 
affected by the local environment.

\acknowledgments

We thank Neal Evans, Fernando Comer\'on, Bruno Mer\'in, Eric Mamajek, and 
Tracy Huard for 
valuable discussions, and the many people who helped get AST/RO ready 
for the 2005 observing season, particularly Jacob Kooi and Craig Kulesa. 
Christina Hammock kept the liquid helium flowing through the winter. We 
thank her, and all the South Pole Station staff, for their work. We thank 
the anonymous referee, whose comments have improved this work. AST/RO 
was supported by the National Science Foundation, under 
NSF OPP ANT-0441756. NFHT was also supported by the University of Exeter 
DVC (Resources) Discretionary Fund and by the European Commission 
(grant MIRG-CT-2006-044961). This work has made use of: NASA's 
Astrophysics Data System; the SIMBAD database operated at CDS, 
Strasbourg; and the {\em Skyview} facility at NASA's Goddard Space 
Flight Center.

{\it Facilities:} \facility{AST/RO}.

\appendix

\section{Frequency Calibration}
\label{app-data}

Because the AOS backends are analog devices, laser mode hopping causes 
shifts in their frequency calibration. A number of such shifts occurred 
over the nine-month period during which these data were taken.

The fundamental frequency calibration for each spectrometer was 
obtained once, by connecting a frequency synthesizer to the IF input 
of the AOS to obtain the AOS channel width and the channel number of 
a fiducial frequency. The frequency scale thus defined was used for 
all data in this paper, but some corrections had to be made. The 
change in channel width caused by a mode-hop is negligible, but the 
entire spectrum is offset by a few channels.

The frequency shifts were tracked with a number of fiducials. Each AOS 
includes a comb generator which is usually used to obtain several 
frequency calibration scans per hour, giving excellent frequency 
tracking. However, the comb generator failed during 2005, so other 
frequency standards had to be used. For the CO 4--3 observations, the 
mesospheric CO 4--3 line is so strong that it can be picked up without 
a switched measurement. So the `sky' spectra, used to estimate the sky 
temperature for calibration, show the line at an antenna velocity close 
to zero. The mesospheric CO abundance has a strong seasonal variation, 
and became so weak at the end of the austral winter (around September) 
that it could no longer be seen in the sky spectra. Finally, repeated 
spectra were taken toward the compact \ion{H}{2} region NGC~3576. This source 
has velocity structure on the scale of the AST/RO beam, so pointing 
uncertainties translate into velocity uncertainties. These three 
fiducials were combined to track the frequency scale. The majority of
\thco\ 2--1 spectra were corrected by 1--3\,\kms, and some were corrected
by up to 5\,\kms. About 90\% of the CO 4--3 spectra were corrected by 
$<1$\,\kms, with the remainder corrected by 3.9\,\kms.

Frequency shifts also show up as inconsistencies in the maps. The 4--3 
observations showed no obvious inconsistencies, but some shifts had to 
be applied to the \thco\ 2--1 data: channel maps of Lupus~I showed that 
data taken in 2005 November had a significant uncorrected frequency 
shift compared to earlier, better-calibrated data. In addition, the 
long-integration spectrum toward one of the reference positions with 
significant 2--1 emission was shifted by about half a channel with 
respect to the map spectra into which it was added. This was not 
corrected, because the facility for combining spectra in COMB only 
handles integer-channel shifts, and the effect of the shift is 
negligible, even in the channel maps.

\section{CO Emission}
\label{app-co}

The CO and \thco\ data in this work are analyzed under the assumption of 
LTE (i.e., the excitation temperature, $T_x$, and gas kinetic temperature,
$T_K$, are the same). For the two transitions considered in this work, 
Equation~(14.46) of \citet{rohlfs} yields
$$T_x=\frac{22.1}{\ln\left(1+\frac{22.1}{T_R^*(4-3)}\right)}$$
for CO 4--3 (neglecting the cosmic microwave background term), and
$$T_x=\frac{10.6}{\ln\left(1+\frac{10.6}{T_R^*(2-1)+0.21}\right)}$$
for \thco\ 2--1. These equations are correct if the transitions are 
optically thick, and completely fill the beam; otherwise they 
underestimate $T_x$. In Lupus, the gas is cold enough ($T_x\la h\nu/k$) that
\trstar\ is significantly different from $T_x$.

The column density in the lower level of the \thco\ 2--1 transition is 
given by
$$N_1^{(13)}=9.69\times 10^{14}\frac{1}{1-\exp(-10.6/T_x)}\int\tau\,dv\,{\rm cm}^{-2}$$
and this can be converted to the total column density of \thco\ by 
correcting for the partition function: $N^{(13)}/N_1^{(13)}$ ranges from 
2.1 to 2.3 for $T_x$ of 7--10\,K.

In LTE, the ratio of \thco\ 2--1 to 1--0 emission depends only on 
temperature. If both transitions are optically thick, the ratio is 
simply the ratio of brightness temperatures at different frequencies 
for a given $T_x$, and will range from 0.7 (at 7\,K) to unity (at high 
temperature). In the optically thin case, this ratio is multiplied by 
the ratio of optical depths, which also depends on $T_x$ via the 
partition function: this ratio ranges from 1 to 2 between 6\,K and 
15\,K, with a high-temperature limit of 3. The 2--1/1--0 line ratio 
should therefore range from 0.7 to about 2 in Lupus.

The ratio of \thco\ 2--1 optical depth to \ceo\ 1--0 optical depth is 
just the ratio of \thco\ 2--1/1--0 optical depths, multiplied by the 
$\mathrm{^{13}CO/C^{18}O}$ abundance ratio: the $\mathrm{^{12}C/^{13}C}$ 
isotope ratio is $62\pm 4$ in the local ISM \citep{93langer}, but the 
double ratio $\mathrm{^{13}CO/C^{18}O}$ is not as well known. Combining 
the solar $\mathrm{^{16}O/^{18}O}$ ratio of 500 \citep{zinner} with the 
local ISM $\mathrm{^{12}C/^{13}C}$ ratio yields 
$\mathrm{^{13}CO/C^{18}O} \sim 8$, but \citet{90langer} point out that 
$\mathrm{^{12}C/^{13}C}$ and $\mathrm{^{16}O/^{18}O}$ should track one 
another, being similarly dependent on star formation history, and so a 
solar ratio \citep[$\mathrm{^{13}CO/C^{18}O} \sim 5.5$,][]{83myers} may 
be more appropriate. Thus the ratio of optical depths should fall in the 
range of 4 to about 16. The ratio of \thco\ 2--1 emission to \ceo\ 1--0, 
however, is complicated by the fact that the \ceo\ transition is likely 
to be fairly optically thin, while the \thco\ transition will have 
moderate to high optical depth. This will tend to reduce the ratio: 
conditions in Lupus are likely to yield emission ratios as low as 2--4.

Below a critical volume density $n_c$, LTE fails ($T_x<T_K$). The 
critical density itself depends on physical conditions: the effective 
spontaneous emission rate $A_{ij}$ is reduced at high optical depth, 
yielding a lower $n_c$, and collisional transition rates are 
temperature-dependent. In the optically thin limit, $n_c$(CO\,4--3) 
varies from $2.9\times 10^4$~\percc\ at 40~K to $4.2\times 10^4$~\percc\ 
at 10~K \citep{jansenthesis}; the critical densities of CO 1--0 and 2--1 
transitions are a few hundred and a few thousand \percc\ \citep{rohlfs}.

\clearpage

\begin{deluxetable}{llllccccl}
\tablewidth{0pt}
\tablecaption{Dark clouds in Lupus
\label{tab-dcs}}
\tablehead{
\colhead{HMSTG\tablenotemark{a}} & \colhead{LM\tablenotemark{b}} & 
\colhead{VMF\tablenotemark{c}} &
\colhead{SL\tablenotemark{d}} &\colhead{$\alpha_{2000}$} & 
\colhead{$\delta_{2000}$} & \colhead{Ref.} & \colhead{YSO\tablenotemark{b}} & \colhead{Other}}
\tablecolumns{9}
\startdata
\cutinhead{Lupus I}
337.9+16.4 & 3379+164   & Lu1     & \nodata  & 15 39 37 & --34 46 30 & LM & Y   & 
FS342\tablenotemark{e} \\
337.6+16.4 & \nodata    & \nodata & \nodata & 15 38 21 & --34 59 17 & HMSTG& 
\nodata & FS341\tablenotemark{e} \\
338.2+16.4 & 3382+164   & Lu4     & \nodata & 15 40 35 & --34 40 19 & LM    & 
N       &  \\
338.8+16.5 & 3388+165-2 & Lu6     & \nodata & 15 42 19 & --33 50 59 & LM    &
N       & B228 \\
           & 3388+165-3 & Lu8     & \nodata & 15 42 40 & --33 52 01 & LM    &
N       & B228 \\
           & 3388+165-4 & Lu7     & SL 12   & 15 42 43 & --34 09 15 & LM    &
Y       & B228 \\
           & 3388+165-5 & B228    & \nodata & 15 43 01 & --34 08 48 & LM    &
Y       & B228 \\
           & 3388+165-6 & Lu9     & \nodata & 15 43 18 & --34 13 30 & LM    &
N       & B228\\
338.7+17.5 & \nodata    & \nodata & \nodata & 15 39 01 & --33 27 38 & HMSTG &
\nodata & B228 \\
339.0+15.8 & \nodata    & Lu12    & \nodata & 15 45 34 & --34 40 53 & VMF   &
\nodata & \\
339.2+16.1 & 3392+161-1 & Lu10    & \nodata & 15 44 54 & --34 17 33 & LM    &
Y       & B228 \\
           & 3392+161-2 & \nodata & \nodata & 15 45 12 & --34 13 21 & LM    &
N       & B228 \\
           & 3392+161-3 & \nodata & \nodata & 15 45 15 & --34 20 43 & LM    &
N       & B228 \\
           & 3392+161-4 & \nodata & SL 13   & 15 45 29 & --34 24 40 & LM    &
N       & FS349\tablenotemark{e}, B228 \\
\nodata    & \nodata    & Lu2     & \nodata & 15 39 56 & --34 42 50 & VMF   &
\nodata & \\
\nodata    & \nodata    & Lu3     & \nodata & 15 40 10 & --33 40 07 & VMF   &
\nodata & B228 \\
\nodata    & \nodata    & Lu5     & \nodata & 15 42 03 & --33 46 33 & VMF   &
\nodata & B228 \\
\cutinhead{Lupus III}
340.7+9.7 & \nodata   & \nodata & \nodata & 16 11 53 & --38 03 48 & HMSTG &
\nodata & Lu~III\,N \\
340.9+9.2 & \nodata   & \nodata & \nodata & 16 13 57 & --38 16 40 & HMSTG &
\nodata & BHR\,134\tablenotemark{f} \\
340.6+9.0 & \nodata   & \nodata & \nodata & 16 14 03 & --38 39 04 & HMSTG &
\nodata & \\
340.2+9.0 & 3402+90-1 & Lu34    & \nodata & 16 11 23 & --39 01 33 & LM    &
PM      & \\
          & 3402+90-2 & Lu36    & \nodata & 16 11 37 & --38 58 21 & LM    &
PM      & \\
          & 3402+90-3 & Lu35    & \nodata & 16 11 45 & --39 01 39 & LM    &
PM      & \\
339.7+9.2 & 3397+92-1 & \nodata & SL 14   & 16 09 42 & --39 09 28 & LM    &
Y       & \\
          & 3397+92-2 & \nodata & \nodata & 16 10 07 & --39 03 47 & LM    &
Y       & \\
          & 3397+92-3 & Lu32    & \nodata & 16 10 23 & --39 10 48 & LM    &
Y       & \\
339.4+9.5 & 3394+95   & Lu30    & \nodata & 16 07 49 & --39 12 04 & LM    &
N       & \\
\nodata   & \nodata   & Lu31    & \nodata & 16 09 08 & --39 03 55 & VMF   &
\nodata & \\
\nodata   & \nodata   & Lu33    & \nodata & 16 10 27 & --39 05 18 & VMF   &
\nodata & \\
\cutinhead{Lupus IV}
336.4+8.2 & 3364+82-1 & Lu23    & \nodata & 16 00 53 & --42 04 08 & LM    &
N       & \\
336.6+7.8 & \nodata   & \nodata & \nodata & 16 02 49 & --42 13 41 & HMSTG &
\nodata & BHR\,120\tablenotemark{f} \\
336.7+8.2 & 3364+82-2 & \nodata & \nodata & 16 01 26 & --41 53 06 & LM    &
N       & \\
336.7+7.8 & \nodata   & \nodata & \nodata & 16 03 15 & --42 06 27 & HMSTG &
\nodata & \\
336.9+8.3 & 3369+83   & Lu25    & SL 7    & 16 02 31 & --41 39 48 & LM    &
N       & \\
336.9+7.8 & \nodata   & Lu26    & \nodata & 16 04 10 & --42 00 40 & VMF   &
\nodata & \\
\nodata   & \nodata   & Lu24    & \nodata & 16 00 18 & --42 03 47 & VMF   &
\nodata & \\
\enddata
\tablenotetext{a}{\citet{86hartley}}
\tablenotetext{b}{\citet{99lm}}
\tablenotetext{c}{\citet{vmf}}
\tablenotetext{d}{\citet{sandy}}
\tablenotetext{e}{\citet{feitzinger}}
\tablenotetext{f}{\citet{bhr1}}
\end{deluxetable}

\begin{deluxetable}{llcccccccccccccccccl}
\tabletypesize{\scriptsize}
\rotate
\tablewidth{0pt}
\tablecaption{Molecular Clouds in Lupus
\label{tab-mcs}}
\tablehead{
\multicolumn{2}{c}{Cloud} &\colhead{$\alpha_{2000}$} 
&\colhead{$\delta_{2000}$} & \multicolumn{3}{c}{CO 4--3} & 
\multicolumn{3}{c}{\thco\ 2--1} & 
\multicolumn{3}{c}{\ceo\ 1--0\tablenotemark{a}} &
\multicolumn{3}{c}{\thco\ 1--0\tablenotemark{b}} &
\multicolumn{3}{c}{\ceo\ 1--0\tablenotemark{b}} 
& \colhead{Notes} \\
\colhead{Hara} & \colhead{VMF} & \colhead{} & \colhead{} &
\colhead{\trstar}&  \colhead{$I$} & \colhead{$\Delta V$} & 
\colhead{\trstar}&  \colhead{$I$} & \colhead{$\Delta V$} & 
\colhead{\trstar}&  \colhead{$I$} & \colhead{$\Delta V$} & 
\colhead{\trstar}&  \colhead{$I$} & \colhead{$\Delta V$} & 
\colhead{\trstar}&  \colhead{$I$} & \colhead{$\Delta V$} & \colhead{}
}
\tablecolumns{20}
\startdata
\cutinhead{Lupus I}
337.6+16.4 & \nodata & 15 38 29 & --35 02 14  & \nodata & \nodata & \nodata & 
2.1 & 4.4 & 2.1 & 0.4     & 0.7     & 1.3     & \nodata & \nodata & \nodata &
\nodata & \nodata & \nodata & \\
338.7+17.5 & \nodata & 15 39 17 & --33 30 06  & 2.3     & 6.2     & 2.8     & 
2.4 & 6.2 & 2.6 & 0.6     & 0.8     & 1.6     & \nodata & \nodata & \nodata &
\nodata & \nodata & \nodata & \\
337.9+16.5 & \nodata & 15 39 24 & --34 45 39  & \nodata & \nodata & \nodata &
1.9 & 3.4 & 1.8 & 2.5     & 1.5     & 0.6     & \nodata & \nodata & \nodata &
\nodata & \nodata & \nodata & \\
338.1+16.7 & \nodata & 15 39 25 & --34 27 33  & \nodata & \nodata & \nodata & 
1.9 & 2.7 & 1.4 & 1.0     & 0.9     & 0.9     & \nodata & \nodata & \nodata &
\nodata & \nodata & \nodata & \\
\nodata    & Lu1     & 15 39 28 & --34 46 22  & \nodata & \nodata & \nodata & 
1.9 & 3.4 & 1.8 & \nodata & \nodata & \nodata &  4.7    & 3.7     & 0.8     &
2.1     &  1.2    & 0.5     & \\
\nodata    & Lu2     & 15 39 56 & --34 42 50  & \nodata & \nodata & \nodata & 
2.0 & 3.0 & 1.5 & \nodata & \nodata & \nodata &  4.6    & 3.3     & 0.7     &
1.9     & 1.0     & 0.5     & \\
\nodata    & Lu3     & 15 40 10 & --33 40 07  & 1.8     & 4.5     &  2.5    & 
1.8 & 3.0 & 1.6 & \nodata & \nodata & \nodata &  5.5    & 9.9     &  1.7    &
0.5     & 0.8     & 1.6     & \\
338.8+17.2 & \nodata & 15 40 14 & --33 38 41  & 2.0     & 5.0     &  2.5    & 
2.4 & 5.7 & 2.4 & 0.5     & 0.8     & 1.3     & \nodata & \nodata & \nodata &
\nodata & \nodata & \nodata & \\
\nodata    & Lu4     & 15 40 32 & --34 39 40  & \nodata & \nodata & \nodata & 
2.3 & 5.0 & 2.2 & \nodata & \nodata & \nodata & 3.3     & 2.9     & 0.8     &
1.2     & 0.6     & 0.5     & \\
\nodata    & Lu5     & 15 42 03 & --33 46 33  & 1.9     & 4.3     & 2.2     & 
1.5 & 4.0 &2.7  & \nodata & \nodata & \nodata & 3.0     & 4.7     & 1.5     &
0.7     & 0.5     & 0.7     & \\
\nodata    & Lu6     & 15 42 04 & --33 50 36  & 1.4     & 2.5     & 1.8     & 
2.1 & 5.2 & 2.4 & \nodata & \nodata & \nodata & 1.9     & 1.9     & 0.9     &
0.5     & 0.4     & 0.8     & \\
\nodata    & Lu7     & 15 42 24 & --34 09 02  & 1.8     & 3.3     & 1.8     &
2.9 & 5.3 & 1.9 & \nodata & \nodata & \nodata & 4.0     & 3.6     & 0.9     &
1.3        & 1.1     & 0.8  & \\
\nodata    & Lu8     & 15 42 35 & --33 52 50  & 1.7     & 4.6     &  2.7    & 
2.5 & 6.0 & 2.4 & \nodata & \nodata & \nodata & 4.5     & 6.2     & 1.3     &
1.1        & 0.8     & 0.7  & \\
338.8+16.5 & \nodata & 15 42 35 & --34 08 58  & 1.6     & 2.8     &  1.8    & 
2.8 & 7.2 & 2.6 & 1.9     & 2.2     & 1.2     & \nodata & \nodata & \nodata &
\nodata & \nodata & \nodata & \\
339.0+16.7 & \nodata & 15 42 48 & --33 53 56  & 1.5     & 4.0     &  2.7    & 
2.3 & 6.6 & 2.8 & 1.4     & 1.6     & 1.1     & \nodata & \nodata & \nodata &
\nodata & \nodata & \nodata & \\
\nodata    & B228    & 15 43 02 & --34 09 06  & 1.8     & 3.3     &  1.9    & 
3.0 & 5.2 & 1.7 & \nodata & \nodata & \nodata & 5.2     & 5.9     & 1.1     &
0.8     & 0.8     & 1.0     & \\  
\nodata    & Lu9     & 15 43 10 & --34 13 50  & 1.6     & 2.3     & 1.5     & 
2.4 & 4.4 & 1.8 & \nodata & \nodata & \nodata & 4.2     & 3.6     & 0.8     &
0.9     & 0.4      & 0.4    & \\
339.1+16.1 & \nodata & 15 44 59 & --34 18 08  & 2.2     &  5.8    & 2.6     & 
3.0 & 9.1 & 3.1 & 1.0     & 1.5     & 1.1     & \nodata & \nodata & \nodata &
\nodata & \nodata & \nodata & \\
\nodata    & Lu10    & 15 45 06 & --34 17 39 & 2.2     &  6.3    & 2.8     & 
2.9 & 8.6 & 3.0 & \nodata & \nodata & \nodata & 3.4    & 3.3     & 0.9     &
0.8     & 0.5     & 0.6     & \\
339.1+15.9 & \nodata & 15 45 30 & --34 25 51 & 3.2     &  4.9    & 1.5     & 
2.9 & 5.9 & 2.0 & 2.3     & 1.5     & 0.7     & \nodata & \nodata & \nodata &
\nodata & \nodata & \nodata & \\
\cutinhead{Lupus III}
\nodata   & lu30    & 16 07 51 & --39 11 12 & \nodata & \nodata & \nodata & 
2.4 & 4.9  & 2.1  & \nodata & \nodata & \nodata & 3.2     & 2.2     & 0.7    &
0.3       & 0.2     & 0.7   & \\ 
339.6+9.3 & \nodata & 16 08 53 & --39 06 26 & 9.0     & 15.5    & 1.7     &
5.6 & 13.1 & 2.3  & 1.9     & 2.4     & 1.2     & \nodata & \nodata & \nodata&
\nodata & \nodata & \nodata & \\
\nodata   & lu31    & 16 09 08 & --39 03 55 & 4.2     & 8.0     & 1.9     & 
3.3 & 8.0  & 2.4  & \nodata & \nodata & \nodata &  1.5    & 1.8     & 1.1    &
0.4     & 0.2     & 0.4     & \\
\nodata   & lu32    & 16 10 19 & --39 12 16 & 2.3     & 3.0     & 1.3     &
2.2 & 5.2  & 2.3  & \nodata & \nodata & \nodata &  4.3    & 3.9     & 0.9    &
0.3     & 0.2     & 0.7     & \\
\nodata   & lu33    & 16 10 27 & --39 05 18 & 2.5     & 3.7     & 1.5     & 
2.1 & 2.9  & 1.4  & \nodata & \nodata & \nodata &  5.1    & 4.1     & 0.8    &
1.2     & 0.7     & 0.6     & \\
\nodata   & lu34    & 16 11 16 & --39 02 38 & \nodata & \nodata & \nodata &
0.9 & 1.9  & 2.1  & \nodata & \nodata & \nodata &  1.0    & 1.0     & 1.0    &
0.5     & 0.2     & 0.4     & \\
\nodata   & lu36    & 16 11 28 & --39 00 21 & \nodata & \nodata & \nodata &
1.3 & 3.2  & 2.5  & \nodata & \nodata & \nodata &  4.6    & 3.7     & 0.8    &
1.2     & 0.7     & 0.5     & \\
\nodata   & lu35    & 16 11 36 & --39 04 18 & \nodata & \nodata & \nodata &
1.4 & 2.7  & 2.0  & \nodata & \nodata & \nodata &  4.4    & 2.8     & 0.6    &
0.6     & 0.5     & 0.7     & \\
340.7+9.7 & \nodata & 16 11 58 & --38 04 23 & 1.5     & 6.0     & 4.1     & 
2.3 & 5.4  & 2.4  & 0.5     & 0.9     & 1.8     & \nodata & \nodata & \nodata&
\nodata & \nodata & \nodata &  Lu~III~N \\
\cutinhead{Lupus IV}
\nodata   & lu24    & 16 00 18 & --42 03 47 & \nodata & \nodata & \nodata &  
1.4 & 2.6 & 1.8 & \nodata & \nodata & \nodata & 4.1     & 2.4     & 0.6     &
0.8     & 0.3     & 0.4     & \\
\nodata   & lu23    & 16 00 57 & --42 04 55 & \nodata & \nodata & \nodata & 
1.6 & 2.9 & 1.8 & \nodata & \nodata & \nodata & 4.2     & 3.8     & 0.9     &
1.2     & 0.7     & 0.6     & \\
336.4+8.2 & \nodata & 16 00 57 & --42 03 16 & \nodata & \nodata & \nodata & 
2.1 & 3.7 & 1.7 & 2.1     & 1.4    & 0.5      & \nodata & \nodata & \nodata &
\nodata & \nodata & \nodata & \\
336.7+8.2 & \nodata & 16 01 46 & --41 52 34 & 3.6     & 5.4     & 1.5     & 
3.4 & 6.5 & 1.9 & 2.0     & 2.0    & 0.8      & \nodata & \nodata & \nodata &
\nodata & \nodata & \nodata & \\
336.9+8.2 & \nodata & 16 02 34 & --41 41 54 & 2.2     & 2.9     & 1.3     & 
2.0 & 3.7 & 1.8 & 1.5     & 1.3    & 0.8      & \nodata & \nodata & \nodata &
\nodata & \nodata & \nodata & \\
\nodata   & lu25    & 16 02 36 & --41 42 26 & 1.2     & 1.2     & 1.0     & 
1.9 & 3.3 & 1.7 & \nodata & \nodata & \nodata & 4.9     & 2.5     & 0.5     &
0.6     & 0.2     & 0.4     & \\
336.7+7.8 & \nodata & 16 03 12 & --42 07 43 & \nodata & \nodata & \nodata &   
1.8 & 3.7 & 2.0 & 0.7    & 0.6     & 0.9      & \nodata & \nodata & \nodata &
\nodata & \nodata & \nodata & \\
336.8+7.9 & \nodata & 16 03 37 & --42 00 55 & \nodata & \nodata & \nodata & 
1.5 & 3.1 & 2.1 & 0.7    & 0.7     & 0.8      & \nodata & \nodata & \nodata &
\nodata & \nodata & \nodata & \\
\nodata   & lu26    & 16 04 10 & --42 00 40 & \nodata & \nodata & \nodata & 
1.4 & 2.7 & 1.8 & \nodata & \nodata & \nodata & 5.2     & 3.8     & 0.7     &
1.1     & 0.6     & 0.5     & \\
\enddata
\tablenotetext{a}{\citet{99hara}}
\tablenotetext{b}{\citet{vmf}}
\end{deluxetable}

\begin{deluxetable}{llcccccccccc}
\tabletypesize{\scriptsize}
\tablewidth{0pt}
\tablecaption{Class 0/I/F YSOs in Lupus
\label{tab-yso}}
\tablehead{
\multicolumn{2}{c}{Designation} & \colhead{Class\tablenotemark{a}} &
\colhead{O/NIR\tablenotemark{b}} &
\colhead{$\alpha_{2000}$} & \colhead{$\delta_{2000}$} & 
\multicolumn{3}{c}{CO 4--3} & \multicolumn{3}{c}{$^{13}$CO 2--1} \\
\colhead{SSTc2d\tablenotemark{a}} & \colhead{Other} & & & & &
\colhead{\trstar} & \colhead{$I$} & \colhead{$\Delta V$} &
\colhead{\trstar} & \colhead{$I$} & \colhead{$\Delta V$} \\}
\tablecolumns{10}
\startdata
\cutinhead{Lupus I}
J154214.6--341026 &                  & I                  & Low & 
15 42 14.6 & --34 10 26 & 1.8     & 2.2     & 1.2     & 2.5 & 4.7 & 1.9 \\
J154301.3--340915 & IRAS 15398--3359 & 0\tablenotemark{c} & \nodata & 
15 43 01.3 & --34 09 15 & 4.3     & 8.2     & 1.9     & 2.8 & 4.8 & 1.7 \\
J154506.3--341738 &                  & F                  & Neb & 
15 45 06.3 & --34 17 38 & 2.1     & 5.9     & 2.8     & 2.9 & 8.6 & 3.0 \\
\cutinhead{Lupus III}
J160703.9--391112 &                  & F                  & IR, Old & 
16 07 03.9 & --39 11 12 & \nodata & \nodata & \nodata & 0.4 & 0.9 & 2.4 \\
J160708.6--391407 &                  & F                  & V, Old & 
16 07 08.6 & --39 14 08 & \nodata & \nodata & \nodata & 0.6 & 1.8 & 3.2 \\
J160829.7--390311 & Sz102            & I                  & Note & 
16 08 29.7 & --39 03 11 & 2.6     & 3.4     & 1.3     & 2.7 & 6.1 & 2.2 \\
J160831.1--385600 &                  & F                  & \nodata & 
16 08 31.1 & --38 56 00 & 0.6     & 1.1     & 1.7     & 0.3 & 0.2 & 0.8 \\
J160846.8--390207 &                  & I                  & \nodata & 
16 08 46.8 & --39 02 07 & 4.2     & 7.8     & 1.9     & 2.7 & 6.0 & 2.3 \\
J160918.1--390453 & Lupus3\,MMS      & 0\tablenotemark{c} & Neb & 
16 09 18.1 & --39 04 53 & 3.5     & 6.2     & 1.8     & 3.3 & 8.0 & 2.4 \\
J160934.2--391513 &                  & F                  & \nodata & 
16 09 34.2 & --39 15 13 & 2.3     & 4.1     & 1.8     & 1.7 & 3.8 & 2.2 \\
J161013.1--384617 &                  & F                  & IR, Old & 
16 10 13.1 & --38 46 17 & \nodata & \nodata & \nodata & 0.4 & 0.6 & 1.4 \\
J161027.4--390230 &                  & F                  & Low & 
16 10 27.4 & --39 02 30 & 2.6     & 4.4     & 1.7     & 1.9 & 4.5 & 2.4 \\
J161204.5--380959 &                  & F                  & V, Low & 
16 12 04.5 & --38 09 59 & \nodata & \nodata & \nodata & 1.2 & 4.3 & 3.7 \\
\cutinhead{Lupus IV}
J160115.6--415235 &                  & F                  & Neb & 
16 01 15.6 & --41 52 35 & 4.5     & 6.4     & 1.4     & 3.6 & 6.2 & 1.7 \\
\enddata
\tablenotetext{a}{\citet{merin}}
\tablenotetext{b}{From \citet{09comeron}: low, below main sequence;
Neb, hidden by nebulosity; IR, suspected IR excess; old, age of order 
100\,Myr; V, suspected veiling; note, discussed individually.}
\tablenotetext{c}{see the text for details}
\end{deluxetable}

\begin{deluxetable}{lccccc}
\tablewidth{0pt}
\tablecaption{Areas mapped and quality of spectra for both transitions
\label{tab-spectra}}
\tablehead{
\colhead{Region} & \colhead{Area(deg$^2$)} & \colhead{Area/cells} 
& \colhead{Spectra} & \colhead{Non-linear\tablenotemark{a}} 
& \colhead{\% non-linear\tablenotemark{a}}}
\tablecolumns{6}
\startdata
\cutinhead{\thco\ 2--1}
Lupus I   & 2.57 & 9258 & 11856 & 0 & 0 \\
Lupus III & 1.87 & 6740 & 8412  & 12 & 0.1 \\
Lupus IV  & 0.92 & 3300 & 3300  & 0 & 0 \\
\sidehead{Total}
Lupus     & 5.36 & 19298 & 23568 & 12 & 0.05 \\
\cutinhead{CO 4--3}
Lupus I     & 0.76 & 10964 & 12067 & 939 & 8 \\
Lupus III N & 0.04 & 625   & 650   & 10  & 2 \\
Lupus III   & 0.17 & 2401  & 3240  & 150 & 5 \\
Lupus IV    & 0.09 & 1225  & 1425  & 15  & 1 \\
\sidehead{Total}
Lupus       & 1.06 & 15215 & 17382 & 1114 & 6 \\
\enddata
\tablenotetext{a}{spectra requiring non-linear baseline subtraction}
\end{deluxetable}

\begin{deluxetable}{lccccc}
\tablewidth{0pt}
\tablecaption{Masses and Line Widths in Lupus
\label{tab-mass}}
\tablehead{
\colhead{Quantity} & \colhead{Lu~I} & \colhead{Lu~III\,N} 
& \colhead{Lu~III} & \colhead{Lu~III\,+} & \colhead{Lu~IV}\\
        & & & & \colhead{Lu~III\,N} & }
\tablecolumns{6}
\startdata
$M/M_\odot$($A_V\ge 2$)     & $280^{+150}_{-40}$ & $40^{+50}_{-10}$ & $110^{+80}_{-20}$ & $150^{+120}_{-20}$ & $80^{+70}_{-10}$ \\
$M/M_\odot$($A_V\ge 3$)     & $150^{+90}_{-40}$ &  $12^{+8}_{-6}$ & $60^{+30}_{-20}$ & $70^{+40}_{-20}$  & $30^{+30}_{-20}$ \\
$\overline{\Delta V}$ (\kms) & 2.3 & 2.7 & 2.2 & 2.4 & 1.5 \\
rms $\Delta V$ (\kms)        & 0.7 & 0.9 & 0.9 & 0.9 & 0.4 \\
\cutinhead{Previous Work}
$M/M_\odot$(\thco\ 1--0)\tablenotemark{a}& 880\tablenotemark{b}   & \nodata & \nodata &  220     &\nodata \\
$\Delta V/$\kms(\thco\ 1--0)\tablenotemark{a}& 1.9   & \nodata & \nodata &  1.7     &\nodata \\
$M/M_\odot$(\thco\ 1--0)\tablenotemark{c}&\nodata & \nodata & \nodata & \nodata &$\la 150$\\
$M/M_\odot$(\ceo\ 1--0)\tablenotemark{d}& 240 & 5 & \nodata & 80 & 160 \\
$\Delta V/$\kms(\ceo\ 1--0)\tablenotemark{d}& 0.6--1.7   & 1.8 & 1.2 &  \nodata     &0.5--0.9 \\
$M/M_\odot$($A_V\ge 2$)\tablenotemark{e}&251 & \nodata & \nodata &  250\tablenotemark{f}  &120\\
\enddata
\tablenotetext{a}{\citet{96tachi}}
\tablenotetext{b}{Covering a larger area than the present work; see 
Section~\ref{sec-results-overview}}
\tablenotetext{c}{\citet{02my}}
\tablenotetext{d}{\citet{99hara}}
\tablenotetext{e}{\citet{c2dresults} and N.~J.~Evans~II (2009, 
private communication)}
\tablenotetext{f}{Assuming a distance of 150\,pc}
\end{deluxetable}

\clearpage

\end{document}